\documentclass[fleqn,usenatbib]{aa}
\usepackage{txfonts}
\usepackage{graphicx}
\usepackage{natbib}
\usepackage{longtable}
\usepackage{subeqnarray}
\usepackage{cases}
\usepackage[colorlinks=true,citecolor=blue]{hyperref}
\usepackage{amsmath}
\usepackage[switch]{lineno} 
\usepackage{multirow}
\usepackage[normalem]{ulem} 
\usepackage{color}
\definecolor{dcoral}{rgb}{0.88, 0.366, 0.366}
\definecolor{fgreen}{rgb}{0.13, 0.55, 0.13}
\newcommand{\orcid}[1]{\href{https://orcid.org/#1}{\includegraphics[width=10pt]{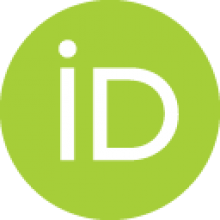}}}
                  
\let\oldequation\equation        
\let\oldendequation\endequation  
\renewenvironment{equation}{\linenomathNonumbers\oldequation}{\oldendequation\endlinenomath} 

\begin{document}
\title{Molecular clouds at the edge of the Galaxy}
\subtitle{II. Physical properties and scaling relations}

\author{C. S. Luo\inst{\ref{inst1},\ref{inst2}}
\and X. D. Tang\inst{\ref{inst1},\ref{inst2},\ref{inst4}}
\and C. Henkel\inst{\ref{inst5},\ref{inst1}}
\and Y. Sun\inst{\ref{inst6}}
\and Y. Gong\inst{\ref{inst6},\ref{inst5}}
\and X. W. Zheng\inst{\ref{inst7}}
\and T. Liu\inst{\ref{inst8}}
\and X. Lu\inst{\ref{inst8}}
\and Y. P. Ao\inst{\ref{inst6}}
\and X. P. Chen\inst{\ref{inst6}}
\and D. L. Li\inst{\ref{inst1},\ref{inst2},\ref{inst4}}
\and Y. X. He\inst{\ref{inst1},\ref{inst2},\ref{inst4}}
\and K. Wang\inst{\ref{inst9}}
\and J. W. Wu\inst{\ref{inst10},\ref{inst2}}
\and J. Esimbek\inst{\ref{inst1},\ref{inst2},\ref{inst4}}
\and J. J. Zhou\inst{\ref{inst1},\ref{inst2},\ref{inst4}}
\and G. Wu\inst{\ref{inst1},\ref{inst2},\ref{inst4}}
\and Y. X. Ma\inst{\ref{inst1},\ref{inst4}}
\and W. A. Baan\inst{\ref{inst1},\ref{inst4}}
\and J. J. Qiu\inst{\ref{inst11}}
\and X. Zhao\inst{\ref{inst1},\ref{inst2}}
\and J. S. Li\inst{\ref{inst1},\ref{inst2}}
\and Q. Zhao\inst{\ref{inst1},\ref{inst2}}
\and L. D. Liu\inst{\ref{inst1},\ref{inst2}}
\and C. Y. Wang\inst{\ref{inst1},\ref{inst2}}
}

\titlerunning{Physical properties and scaling relations in the Galactic edge clouds}
\authorrunning{Luo et al.}

\institute{
State Key Laboratory of Radio Astronomy and Technology, Xinjiang Astronomical Observatory, CAS, Urumqi 830011, PR China\label{inst1}\\
\email{tangxindi@xao.ac.cn}
\and University of Chinese Academy of Sciences, Beijing 100080, PR China \label{inst2}
\and Xinjiang Key Laboratory of Radio Astrophysics, Urumqi 830011, PR China \label{inst4}
\and Max-Planck-Institut f\"{u}r Radioastronomie, Auf dem H\"{u}gel 69, 53121 Bonn, Germany \label{inst5}
\and Purple Mountain Observatory, Chinese Academy of Sciences, Nanjing 210008, PR China \label{inst6}
\and School of Astronomy and Space Science, Nanjing University, Nanjing 210093, PR China \label{inst7}
\and Shanghai Astronomical Observatory, Chinese Academy of Sciences, 80 Nandan Road, Shanghai 200030, PR China \label{inst8}
\and Kavli Institute for Astronomy and Astrophysics, Peking University, Beijing 100871, PR China \label{inst9}
\and National Astronomical Observatories, Chinese Academy of Sciences, Beijing 100101, PR China \label{inst10}
\and School of Mathmatics and Physics, Jinggangshan University, Ji’an 343009, PR China \label{inst11}
}
\date{Received February 02, 2026, Accepted April 18, 2026}

\abstract
{The outer Galaxy presents an optimal setting for investigating molecular clouds and star formation in environments with 
low metallicity. A total of 72 Galactic edge clouds were surveyed using the CO\,(2--1) line with the IRAM\,30\,m telescope, 
leading to the identification of 112 CO clumps within molecular clouds with linear resolutions of 0.5--0.9\,pc. 
Parameters such as size, mass, surface density, and velocity dispersion of these CO clumps, derived from CO\,(2--1) observations, 
exhibit ranges of 0.6--3.4\,pc, 34--8250\,M$_\odot$, 12--1025\,M$_{\odot}$\,pc$^{-2}$, and 0.3--1.7\,km\,s$^{-1}$, respectively.
Over the Galactocentric distance range of 14--23\,kpc, no systematic variations are found in these parameters.
The velocity dispersion-size relationship of the Galactic edge clumps is modeled as 
$\sigma_{\rm v}$\,=\,0.69($\pm$0.03)$R_{\rm eff}^{0.36(\pm0.10)}$, indicating that turbulence is present within the Galactic 
edge clumps, akin to observations in the inner Galactic disk clouds. 
Furthermore, the luminous mass-size relation of the Galactic edge clumps is described by 
$M_{\rm lum}$\,=\,196($\pm$17)$R_{\rm eff}^{\,2.18\,(\pm0.26)}$, suggesting the average column density remains almost 
constant for clouds of different sizes. The virial parameters range from 0.6 to 15.3, with a median value of 2.8\,$\pm$\,0.6, 
suggesting that most clumps are gravitationally unbound. Furthermore, the virial parameters of our Galactic edge clumps 
show a decreasing trend with increasing Galactocentric distances, described by an exponential relation 
$\alpha_{\rm vir}$\,=\,33.0($\pm$\,10.4)\,e$^{-R_{\rm g}/6.7(\pm0.9)}$, consistent with previous results. 
}

\maketitle

\section{Introduction} 
\label{sect:Introduction} 
Outskirts of the Galaxy, known as low-density and metal-poor environments, provide an ideal laboratory to study molecular 
clouds and star formation \citep{Kobayashi2008, Yasui2008}. However, molecular clouds at the edge of the Galaxy exhibit a
spatially extended yet sparse distribution across the sky, owing to their remote locations. The pressure exerted by the 
inter-cloud medium is expected to be low compared to that in the inner Galaxy. External triggers for star formation are 
either absent or only weakly present, with spiral arms and supernova remnants being less significant \citep{Brand1995}. 
Observational studies of molecular gas in the outer Galaxy have been carried out and star formation activity has been found
to be present (e.g., \citealt{Fich1984,Mead1988,Wouterloot1988,Wouterloot1989,Digel1994,Heyer2001,Du2016,Su2016,Sun2020,Sun2024a,Sun2024b,Ma2021,Ma2022,Braine2023,Urquhart2024,Cheng2025,Lin2025,Luo2025}). This includes star-forming molecular clouds 
(e.g., \citealp{Gong2016,Izumi2017,Izumi2022,Izumi2024,Ikeda2025}), stellar clusters 
(e.g., \citealt{Yasui2006,Yasui2008,Kobayashi2008,Izumi2014}), \ion{H}{II} regions 
(e.g., \citealt{deGeus1993,Rudolph1996,Anderson2015}), and maser lines (e.g., \citealt{Armentrout2017,Sun2018a}). 
Nevertheless, the Galactic edge clouds have not been subjected to the same level of detailed observational studies 
as nearby clouds. 

Distinct boundaries and hierarchical structures have been observed in molecular clouds 
(e.g., \citealt{Goodman1998,McKee2007,Rosolowsky2008,Yuan2021,Shen2024a}). 
Molecular cloud structures can be categorized into clouds ($\sim$\,10\,pc), clumps ($\sim$\,1\,pc), and cores ($\sim$\,0.1\,pc). 
These entities are successively smaller in size and higher in density \citep{Williams2000,Kauffmann2013}. 
The majority of stars originate from the dense regions within molecular clouds, where gravity is sufficiently strong to 
induce gas collapse \citep{Bergin2007}. 
The clump properties reflect the initial conditions for stellar cluster formation \citep{Mok2021}. 
Increasing evidence suggests that the physical properties of the interstellar medium influence the star formation rate, 
the spatial distribution and key characteristics of subsequent generations of stars, including elemental composition
and initial mass function 
(e.g., \citealt{Paumard2006,Kennicutt1998a,Kennicutt1998b,Klessen2007,Papadopoulos2011,Zhang2018,Tang2019,Gong2025}).
It is crucial to understand the clump properties and the factors affecting them. The typical masses, sizes, and volume 
densities of clumps in cold dark clouds are 50--500\,M$_\odot$, 0.3--3.0\,pc, and $10^3$--$10^4$\,cm$^{-3}$, 
respectively \citep{Bergin2007}. So far, the physical properties of clumps in Galactic edge clouds remain poorly constrained. 
Previous CO surveys lack the angular resolution required to resolve individual molecular clumps, hindering a detailed 
characterization of their physical properties. Obtaining further high-angular-resolution CO observations is instrumental
in elucidating the clump's properties at the Galactic edge. 

The three scaling relations proposed by \cite{Larson1981} serve as fundamental observational constraints on the dynamics 
of molecular clouds. \cite{Larson1981} established a power-law relation between velocity dispersion ($\sigma_{\rm v}$) 
and cloud size ($L$) with an index of 0.38, an inverse correlation between the mean density of the cloud ($n$(H$_2$)) 
and $L$, and the relation 2G$M/\sigma_{\rm v}^2L$\,$\sim$\,1, where G and $M$ denote the gravitational constant and 
the cloud mass, respectively. These relations are traditionally interpreted sequentially to suggest that molecular 
clouds are turbulent structures, the mass surface density ($\Sigma$) remains relatively constant among molecular clouds, 
and there exists an equipartition between gravitational and kinetic energy densities \citep{Heyer2015}. Notably, 
these relations are interdependent, where the validity of any two expressions algebraically implies the third. 
This interconnection is encapsulated in the unified expression equating the virial mass to the cloud mass \citep{Heyer2009,Heyer2015}, 
\begin{equation}
\sigma_{\rm v}\,=\,(\pi {\rm G}/5)^{1/2}R^{1/2}\Sigma^{1/2}\,,
\label{eq:size-mass-linewidth} 
\end{equation}
where $R$\,=\,$L$/2 represents the cloud radius. This expression implies a scaling exponent of 1/2 for the velocity 
dispersion–size relation for constant mass surface density, diverging from the original value of $\sim$\,0.38 derived 
by \cite{Larson1981}.

Numerous surveys utilizing multiple molecular tracers across diverse scales in both our Galaxy and external galaxies 
have significantly contributed to the advancement and enhancement of these scaling relations. These surveys offer a 
consistent dataset with extensive samples of molecular clouds encompassing a wide range of environmental conditions. 
The velocity dispersion-size relation (hereafter $\sigma_v$--$R$ relation) has been investigated in different environments, 
including the Milky Way
(e.g., \citealt{Larson1981,Solomon1987,Heyer2001,Heyer2009,Shetty2012,Heyer2015,Rice2016,Sun2017,Sun2024a,Miville-Deschenes2017,Ma2021,Feng2024,Luo2024a,Luo2024b}) and nearby galaxies 
(e.g., \citealt{Wong2017,Wong2019,Wong2022,Sun2018c,ONeill2022b,Saldano2023,Finn2024a,Krahm2024}). 
Observational results suggest that the power-law index predominantly varies from 1/3 to 1/2 between velocity dispersion
and size (e.g., \citealt{Liu2012,Colombo2014,Maeda2020}). 
Giant molecular clouds (GMCs) are typically defined as molecular gas structures with masses exceeding 10$^5$\,M$_\odot$ 
and spatial extents of $\sim$\,10--100\,pc \citep{Blitz1993}. Within GMCs, the power-law index approximating 1/2 
is indicative of supersonic turbulence, while the power-law index of 1/3 suggests the presence of subsonic 
turbulence \citep{McKee2007,Indebetouw2013}.

The relation between mass and size (hereafter $M$--$R$ relation) in GMCs also exhibits a power-law scaling of 
$M$\,$\propto$\,$L^q$. Previous studies indicate that the $q$ value typically falls within the range of 1 to 3 
(e.g., \citealt{Larson1981,Solomon1987,Shetty2010,Kauffmann2010a,Kauffmann2010b,Kauffmann2010c,Miville-Deschenes2017,Faesi2018,Traficante2018,Ballesteros-Paredes2019,Lada2020,Li2020,Ma2021,Xing2022}). 
$q$\,=\,2 implies that molecular clouds maintain a nearly constant average column density or surface density as a function of size. 
Nevertheless, the $M$--$R$ relation within individual clouds can not be adequately represented by a single power-law, 
as $q$ demonstrates variation with radius \citep{Kauffmann2010a,Kauffmann2010b}. The empirical threshold of 
the $M$--$R$ relation for massive star formation generally adheres to $M(R)$\,>\,580M$_{\odot}$($R$/pc)$^{1.33}$ \citep{Kauffmann2010c}.

As mentioned above, while numerous studies have been conducted on the physical properties of molecular clouds, 
observational biases are unavoidable owing to limited spatial resolution and sensitivity. 
To mitigate these biases, observations spanning a larger dynamic range are required. 
The outer regions of the Galaxy may be younger than its inner counterparts, as evidenced by the presence of 
young stellar populations \citep{Martig2016}. The outer Galaxy exhibits characteristics akin to those of dwarf 
galaxies and the Milky Way in the early phase of formation \citep{Izumi2024}. 
Previous observations suggest that about half of the molecular gas is CO-dark in the solar neighborhood (e.g., \citealt{Paradis2012,Pineda2013,Chen2015}). This contribution is expected to be even more pronounced in the low metallicity environment of the Galactic edge clouds \citep{Pineda2013,Langer2014,Luo2024c}. 
The physical properties and scaling relations of molecular clouds have been studied in the outer Galaxy 
(e.g., \citealt{Mead1988,Brand1995,Wouterloot1995,May1997,Brand2001,Heyer2001,Su2016,Sun2017,Sun2024a,Ma2021,Urquhart2024,Yuan2025}). 
However, such clump-scale studies at the edge of the Galaxy remain limited. Recently, a large number of 
molecular clouds in the outer Galaxy have been identified by the Milky Way Imaging Scroll Painting (MWISP) project, 
an unbiased CO survey of the northern Galactic plane with an angular resolution of $\sim1'$
\citep{Sun2015a,Sun2024a,Sun2024b,Su2019,Yang2026}. While MWISP provides an excellent census of outer-Galaxy 
molecular clouds with Galactocentric distances up to $\sim$26\,kpc, 
its angular resolution is insufficient to resolve individual molecular clumps at the edge of the Galaxy. 
Mapping these molecular clouds with higher sensitivity and angular resolution is therefore essential for characterizing
clump-scale physical properties and star formation at the edge of the Galaxy. 

The present study aims to observe Galactic edge clouds at high resolution ($\sim$\,11$^{\prime\prime}$) in an 
effort to investigate the physical properties of molecular clumps and to explain their equilibrium state by scaling relations.  
In Sects.\,\ref{sect:Observations and data reduction} and \ref{sect:Results}, we introduce our observations, 
data reduction, and describe the main results. The discussion of the clump's dynamical state is presented in 
Sect.\,\ref{sect:discussion}. Our main conclusions are summarized in Sect.\,\ref{sect:summary}. 
This paper is part of the "Molecular Clouds at the Edge of the Galaxy" project, which focuses on the investigation 
of the physical and chemical properties of molecular clouds and star formation located in the outskirts of the Milky Way. 

\setcounter{figure}{0}
\begin{figure*}[h]
\centering
\includegraphics[width=1.0\textwidth]{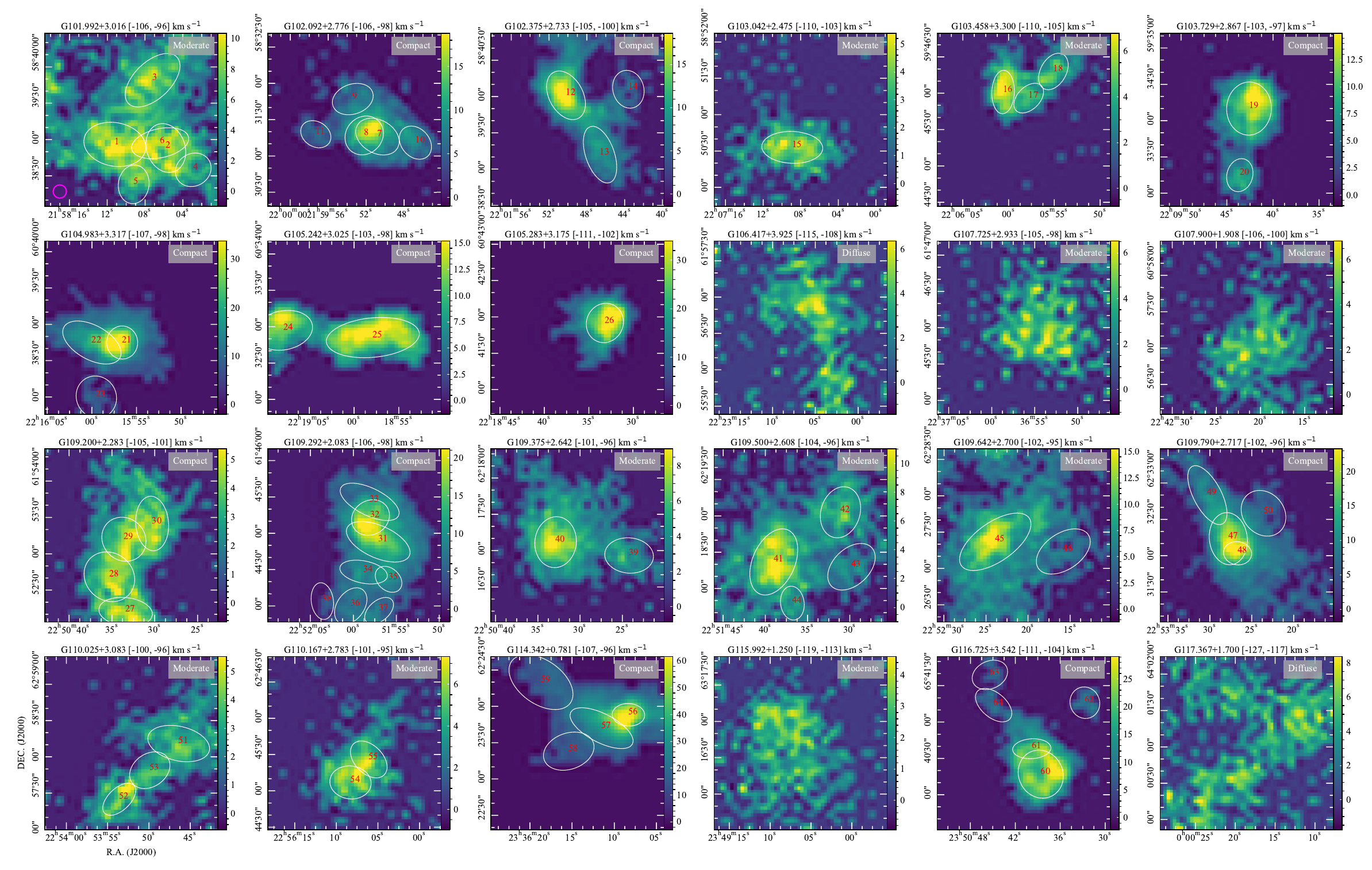}
\caption{CO\,(2--1) velocity-integrated intensity maps of 72 Galactic edge clouds. 
The color background shows the integrated main beam brightness temperature, $\int T_{\rm MB}\,dv$. 
For each panel, the source name and the velocity-integration range are indicated at the top, and the color bar 
(in units of K\,km\,s$^{-1}$) is shown on the right. Source names follow those listed in Table\,C.1 of \citet{Luo2025}. 
The molecular cloud type (compact, intermediate, or diffuse), based on visual classification, is indicated in 
the upper-right corner of each panel. Ellipses denote the identified and fitted CO clumps
(see Sect.\,\ref{sect:Overview}), labelled from 1 to 112 and corresponding to the indices in Table\,\ref{table:parameters}. 
The magenta circle in the lower-left corner of the first panel indicates the beam size of the IRAM\,30\,m CO\,(2--1)
observations. Refer to Fig.\,\ref{fig:map2} for the velocity-integrated intensity maps of other targets.}
\label{fig:map} 
\end{figure*}

\section{Observations and data reduction}
\label{sect:Observations and data reduction}
\subsection{Observations}
\label{sect:Observations}
An extension of the Scutum-Centaurus arm has been detected beyond the Outer arm \citep{Dame2011,Sun2015a,Sun2024b}.
\citet{Sun2015a} catalogued 72 molecular clouds located in the second Galactic quadrant ($100^\circ<l<150^\circ$) of 
this arm, which is situated at the edge of the Galaxy ($R_{\rm g}$\,$\sim$\,14\,--22\,kpc, see also Fig.\,1 in \citealt{Luo2025}). 
The kinematic distances of these Galactic edge clouds were determined by \cite{Sun2015a} using the Galactic rotation 
curve from \cite{Reid2014}. In the present study, we recalculated these distances employing the updated Galactic rotation 
curve model from \cite{Reid2019}. The derived results are listed in Table\,\ref{table:parameters}. The mean Galactocentric distance of these Galactic edge clouds is $\sim$\,16.7\,kpc, 
corresponding to a metallicity of $\sim$\,0.3\,$Z_{\odot}$ ($Z_{\odot}$ signifies the Solar metallicity; \citealt{Mendez-Delgado2022}). 
Utilizing the IRAM\,30\,m telescope\footnote{\tiny Based on observations obtained with the IRAM\,30\,m telescope. 
IRAM is supported by INSU/CNRS (France), MPG (Germany), and IGN (Spain).}, high-sensitivity, high-resolution maps 
of the CO\,(2--1) line were obtained towards these 72 Galactic edge clouds. Maps of size
$\sim$\,100$^{\prime\prime}$\,$\times$\,100$^{\prime\prime}$ are generated by the On-The-Fly (OTF) observing mode. 
The Half Power Beam Width (HPBW) is $\sim$\,11$^{\prime\prime}$ at a central frequency of $\sim$\,230\,GHz. 
The heliocentric distances ($D_{\rm s}$) of these Galactic edge clouds vary between approximately 9 and 16\,kpc, 
with a median of $\sim$\,10\,kpc. Correspondingly, the spatial linear scales range from 0.5 to 0.9\,pc at a beam 
size of 11$^{\prime\prime}$, with an average of $\sim$\,0.6\,pc. The typical root-mean-square (RMS) noise levels 
are $\sim$\,0.3--0.5\,K on a main beam brightness temperature ($T_{\rm mb}$) scale at a channel width of $\sim$\,0.25\,km\,s$^{-1}$. 
Further details regarding the targets and observations can be found in \citet{Luo2025}. 

\subsection{Data reduction}
\label{sect:Data reduction}
The spectral lines observed were processed using the GILDAS\footnote{\tiny \url{http://www.iram.fr/IRAMFR/GILDAS}} 
and {\tt Python}\footnote{\tiny \url{https://www.python.org/}} packages. 
The signal-to-noise ratio (S/N) was improved by eliminating low-quality CO\,(2--1) data and merging dual-polarization 
spectral lines. The $T_{\rm mb}$ values for the CO\,(2--1) have been calibrated adopting a 
main beam efficiency ($B_{\rm eff}$\,$\sim$\,0.59) and a forward hemisphere efficiency ($F_{\rm eff}$\,$\sim$\,0.92), 
using the expression $T_{\rm mb}$\,=\,($F_{\rm eff}/B_{\rm eff}$)$T_{\rm A}^{*}$. 
The raw data were subsequently re-gridded into pixels measuring 5.5$^{\prime\prime}$\,$\times$\,5.5$^{\prime\prime}$. 

\begin{figure*}[t]
\centering
\includegraphics[width=0.95\textwidth]{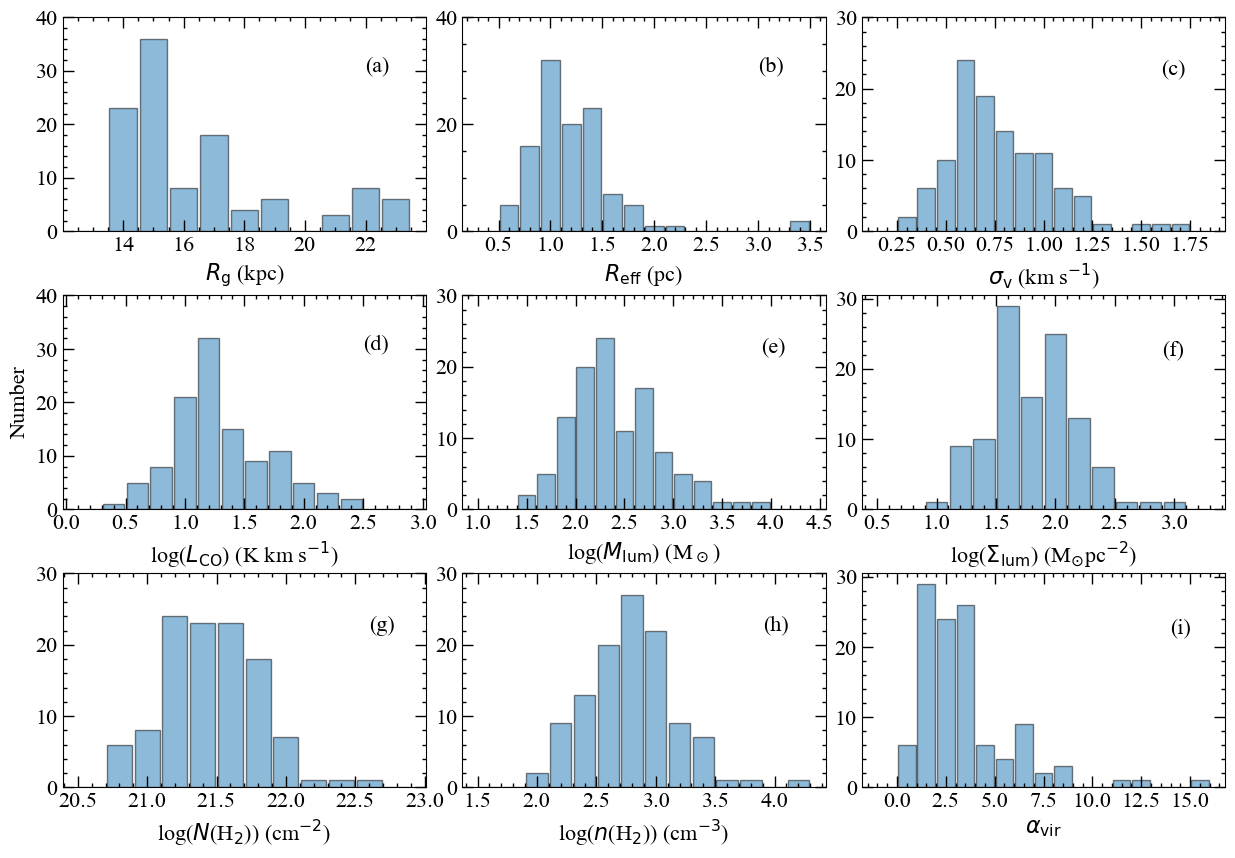}
\caption{Histogram of CO clump's Galactocentric distances (a), effective radii (b), velocity dispersions (c), 
CO luminosities (d), luminous masses (e), surface densities (f), H$_2$ column densities (g), mean volume densities (h), 
and virial parameters (i). 
} 
\label{fig:pra_hist} 
\end{figure*}

\section{Results}
\label{sect:Results}
\subsection{Overview}
\label{sect:Overview} 
The CO\,(2--1) line has been detected in all 72 observed molecular clouds at the edge of the Milky Way. 
The integrated intensity distributions are presented in Figs.\,\ref{fig:map} and \ref{fig:map2}. 
Morphological structures of these clouds have been resolved by the CO\,(2--1) data at linear resolutions of 0.5--0.9\,pc. 
Overall, 25 clouds exhibit compact morphologies (e.g., G139.116$-$1.475), 25 show diffuse structures (e.g., G145.808$-$1.817), 
and 22 display intermediate morphologies between these two categories (e.g., G101.992$+$3.016; see Fig.\,\ref{fig:map} 
and Table\,C.1 of \citealt{Luo2025}). Typical examples of these three cloud structure classes are also illustrated 
in Fig.\,A.1 of \citet{Luo2025}. In compact clouds, 81 clumps have been identified, whereas in intermediate clouds 
and diffuse clouds, the numbers stand at 28 and 3 (see Sect.\,\ref{sect:Clump identification}), respectively. 
The clumps predominantly reside in compact clouds, with their presence diminishing in intermediate and diffuse clouds. 
Clouds exhibiting clumpy morphology demonstrate a higher degree of star formation \citep{Neralwar2022,Luo2025}. 
The morphology of molecular clouds at the edge of the Galaxy exhibits similarities to those observed in the 
less extreme outer Galaxy, notably characterized by filamentary, clumpy, and extended structures \citep{Yuan2021}. 

\begin{table}[t]
\scriptsize
\caption{Statistical parameters of identified CO clumps.}
\centering
\begin{tabular}
{lcccc}
\hline\hline
Properties &Min &Max &Mean &Median\\
\hline
$R_{\rm g}$ (kpc)                           &14.2\,$\pm$\,0.1 &22.9\,$\pm$\,0.7  &16.7\,$\pm$\,0.3   &15.4\,$\pm$\,0.3 \\
$R_{\rm eff}$ (pc)                          &0.6 \,$\pm$\,0.1 &3.4 \,$\pm$\,0.2  &1.2 \,$\pm$\,0.1   &1.2 \,$\pm$\,0.1 \\
$\sigma_{\rm v}$ (km\,s$^{-1}$)             &0.3 \,$\pm$\,0.1 &1.7 \,$\pm$\,0.1  &0.8 \,$\pm$\,0.1   &0.7 \,$\pm$\,0.1 \\
$L_{\rm CO}$ (K\,km\,s$^{-1}$\,pc$^{2}$)    &3.0 \,$\pm$\,0.4 &285 \,$\pm$\,31   &32.5\,$\pm$\,4.1   &16.8\,$\pm$\,2.3 \\
$M_{\rm lum}$ (M$_\odot$)                   &34.2\,$\pm$\,6.5 &8250\,$\pm$\,993  &518 \,$\pm$\,82    &216 \,$\pm$\,40  \\
$\Sigma_{\rm lum}$ (M$_{\odot}$\,pc$^{-2}$) &11.7\,$\pm$\,3.0 &1025\,$\pm$\,261  &92.9\,$\pm$\,18.7  &62.6\,$\pm$\,13.0\\
$N$(H$_2$) ($10^{21}$\,cm$^{-2}$)           &0.5 \,$\pm$\,0.2 &47.7\,$\pm$\,12.1 &4.3 \,$\pm$\,0.9   &2.9 \,$\pm$\,0.6 \\
$n$(H$_2$) ($10^{2}$cm$^{-3}$)              &1.0 \,$\pm$\,0.3 &161 \,$\pm$\,30   &9.8 \,$\pm$\,1.8   &6.3 \,$\pm$\,1.1 \\
$\alpha_{\rm vir}$                          &0.6 \,$\pm$\,0.1 &15.3\,$\pm$\,3.5  &3.4 \,$\pm$\,0.8   &2.8 \,$\pm$\,0.6 \\
\hline
\end{tabular}
\label{table:statistics of physical parameters}
\end{table}

\begin{figure*}[t]
\centering
\includegraphics[width=0.95\textwidth]{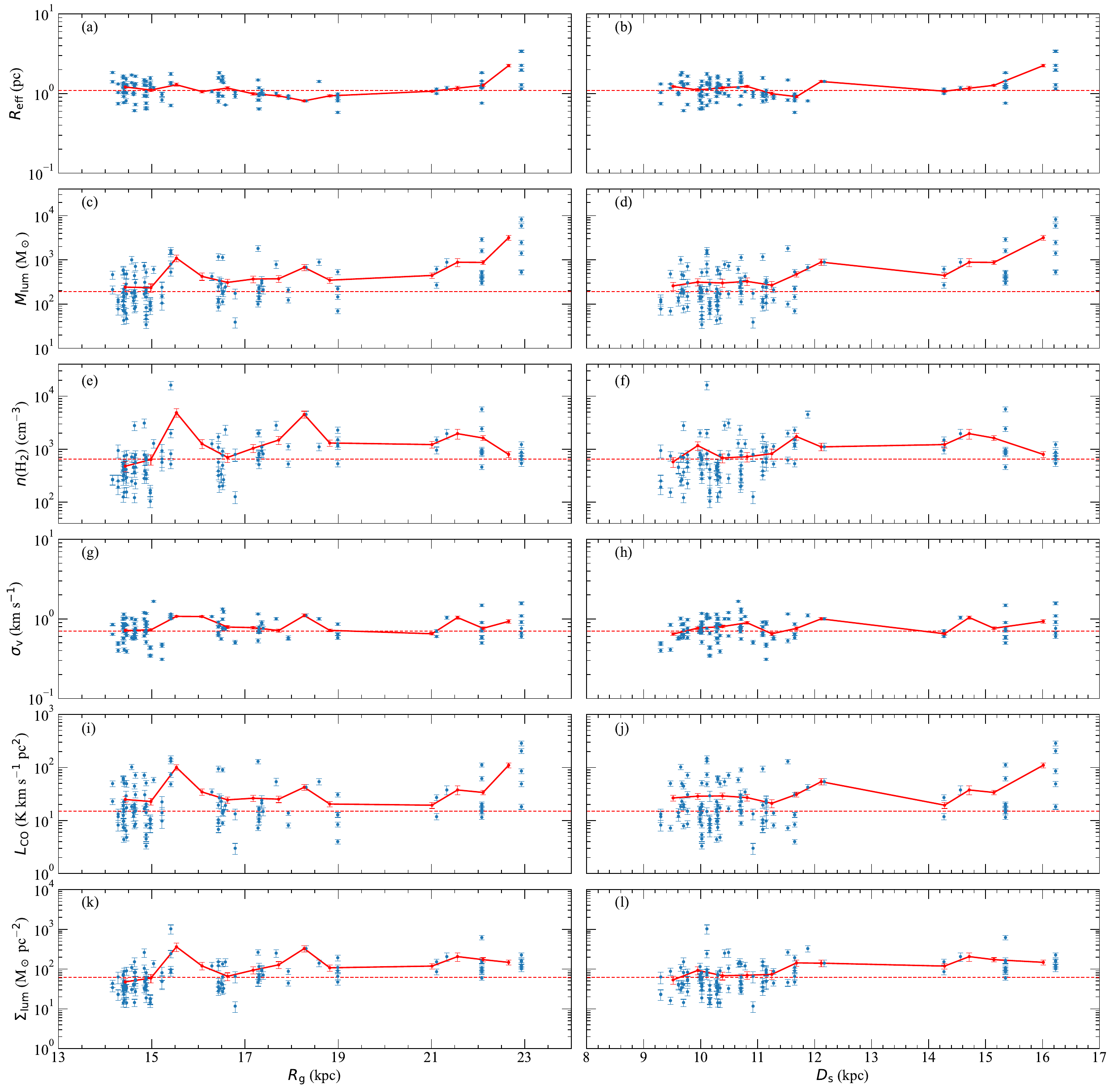}
\caption{Variation in CO clump's properties with Galactocentric (\emph{left column}) and heliocentric distances 
(\emph{right column}): effective radius (a--b), luminous mass (c--d), volume density (e--f), velocity dispersion (g--h), 
CO luminosity (i--j), and surface density (k--l). The solid red lines connecting the data points represent the mean 
value within a bin, with each bin referring to 0.5\,kpc. The horizontal dashed red line in each panel denotes a median 
value (also see Table\,\ref{table:statistics of physical parameters}). 
} 
\label{fig:pra_dis} 
\end{figure*}

\subsection{Clump identification}
\label{sect:Clump identification} 
Clumps within our Galactic edge clouds are identified using the Python package, 
{\tt quickclump}\footnote{\tiny \url{https://github.com/vojtech-sidorin/quickclump/}} \citep{Sidorin2017}. 
This algorithm characterizes molecular clumps in a manner comparable to other segmentation methods, such as CLUMPFIND
\citep{Williams1994}. The {\tt quickclump} method has been extensively employed in segmenting molecular line emissions 
(e.g., \citealt{Indebetouw2020,Finn2021,Finn2022,Finn2024a,ONeill2022b,Krahm2024}). Considering observational Flexible 
Image Transport System (FITS) cubes, encompassing the three-dimensional position-position-velocity (3D-PPV) space, 
the {\tt quickclump} algorithm requires four input parameters. The parameter $T_{\rm cutoff}$ denotes the minimum intensity level for clump identification. $dT_{\rm leaf}$ is the minimum intensity difference between adjacent emission peaks that allows 
them to be identified as distinct clumps. $I_{\rm minpk}$ is an added 
parameter\footnote{\tiny \url{https://github.com/indebetouw/quickclump}} (see below), the minimum peak intensity required for 
a valid clump \citep{Indebetouw2020}. $N_{\rm pix}$ represents the minimum number of voxels necessary for a clump. 

In this study, molecular clumps are identified using the following parameters: $T_{\rm cutoff}$\,=\,3.0\,$\sigma_{\rm rms}$, 
$dT_{\rm leaf}$\,=\,2.0\,$\sigma_{\rm rms}$, $I_{\rm minpk}$\,=\,5.0\,$\sigma_{\rm rms}$, $N_{\rm pix}$\,=\,2 beams, 
where $\sigma_{\rm rms}$ is the RMS noise level of the spectral data extracted from the respective FITS file. 
The parameter $T_{\rm cutoff}$ was set to 3\,$\sigma_{\rm rms}$ to ensure that only statistically significant emissions are included, while $N_{\rm pix}$ was chosen to cover at least two beams to avoid spurious detections caused by noise fluctuations. 
$dT_{\rm leaf}$ was set to 2\,$\sigma_{\rm rms}$ to balance the identification of distinct clumps with the avoidance of over-segmentation, and $I_{\rm minpk}$ was set to 5\,$\sigma_{\rm rms}$ to ensure a high signal-to-noise ratio for the detected clumps. 
To assess the robustness of clump identification, small perturbations of 10\% were applied to the segmentation parameters, resulting in variations of fewer than six in the number of identified clumps. 
During the identification of clumps, map edge regions with significant noise were excluded, resulting in an area of 
26\,$\times$\,26 pixels, equivalent to 143$^{\prime\prime}$\,$\times$\,143$^{\prime\prime}$.
The algorithm automatically identified 165 clumps, of which 53 were manually removed because they were either 
located at the edge of a map or smaller than the beam size. Fig.\,\ref{fig:channel maps} shows examples of clumps 
identified in the 3D-PPV space for different velocity channels in the G145.208-0.392 molecular cloud. 
Consequently, a sample of 112 CO clumps has been identified in the 72 Galactic edge clouds (see Figs.\,\ref{fig:map} and \ref{fig:map2}). 

\subsection{Estimation of physical parameters} 
\label{sect:Estimation of Physical Parameters} 
The local standard of rest velocity ($V_{\rm LSR}$) and velocity dispersion ($\sigma$) are obtained by 
fitting a Gaussian profile\footnote{\tiny \url{https://scipy.org/}} to the intensity-weighted mean spectrum of each clump. 
The velocity dispersion ($\sigma_{\rm v}$) is subsequently derived by deconvolving $\sigma$ with the channel width 
($\Delta_{\rm v}$, 0.25\,km\,s$^{-1}$ in our observation), 
\begin{equation}
\sigma_{\rm v} = \sqrt{\sigma^2-(\Delta_{\rm v}/\sqrt{8\ln2})^2} \,. 
\label{Standard deviation of velocity}
\end{equation}
Uncertainties in $\sigma_{\rm v}$ arise from the fitting procedure and the deconvolution process. 
All CO clumps exhibit a single velocity component. 
The $\sigma_{\rm v}$ ranges from 0.3 to 1.7\,km\,s$^{-1}$, with a median value of 0.7\,$\pm$\,0.1\,km\,s$^{-1}$ 
(see Table\,\ref{table:statistics of physical parameters}). 
The line width, defined as the full width at half maximum (FWHM) of CO\,(2--1), is related to $\sigma_{\rm v}$ 
by ${\rm FWHM}$\,=\,$\sqrt{8\ln2}$\,$\sigma_{\rm v}$. 

The effective radius ($R_{\rm eff}$) is derived by fitting an ellipse to each CO clump. 
The half-width at half-maximum for both the major ($w_{\rm maj}$) and minor ($w_{\rm min}$) axes\footnote{\tiny The $w_{\rm maj}$ 
and $w_{\rm min}$ are given in pixels and need to be converted to arcsec, followed by a conversion to pc scale.} 
is converted to an equivalent Gaussian dispersion 
\begin{equation}
\sigma_{\rm R} = 2\sqrt{w_{\rm maj} w_{\rm min}}/\sqrt{8\ln2} \,. 
\label{Standard deviation of radius}
\end{equation}
To determine the effective radius of a clump, the measured extent is deconvolved from the beam size, 
\begin{equation}
R_{\rm eff} = \eta \sqrt{\sigma_{\rm R}^2-(\theta_{\rm beam}/2)^2} \,,
\label{effective radius}
\end{equation}
where $\theta_{\rm beam}$ is the beam size in pc. 
The $\eta$ value of 1.91 is adopted to convert the Gaussian dispersion into an effective physical radius corresponding 
to a uniform-density sphere, in accordance with previous studies
(e.g., \citealt{Solomon1987,Rosolowsky2006,Wong2022,ONeill2022b,Finn2024a,Krahm2024}). 
For these CO clumps, the sizes range from 0.6 to 3.4\,pc, with a median of 1.2\,pc (see Table\,\ref{table:statistics of physical parameters}). 
The statistical size distribution of these clumps does not distinctly adhere to power-law forms (see Fig.\,\ref{fig:pra_hist}). 

We assume that the CO\,$J$=2--1 and $J$=1--0\footnote{\tiny CO\,$J$=1--0 data were obtained as part of the 
MWISP project using the Delingha 13.7\,m telescope with a beam size of $\sim$\,50$^{\prime\prime}$ \citep{Sun2015a}.} 
lines may trace similar components of molecular gas (e.g., \citealt{Luo2025}). 
For clumps extracted from the CO\,(2--1) data, the CO luminosities ($L_{\rm CO}$) are determined by 
\begin{equation}
L_{\rm CO} = A_{\rm pix} \frac{\Sigma I_{\rm CO(2-1)}}{R_{21}} \,,
\label{eq:luminosities}
\end{equation}
where $A_{\rm pix}$ denotes the area of a pixel, measured in pc$^2$. 
$\Sigma I_{\rm CO(2-1)}$ signifies the CO\,(2--1) integrated intensities in K\,km\,s$^{-1}$ summed over all pixels in a clump. 
$R_{21}$ represents the CO\,$J$=2--1/1--0 integrated line brightness ratio. 
In each Galactic edge cloud, all clumps are assumed to share a common $R_{21}$ value, as detailed in \citet{Luo2025}. 
The $R_{21}$ ratios span from 0.30 to 3.04, with a mean of 0.97\,$\pm$\,0.12 and a median of 0.81\,$\pm$\,0.10. 
Consequently, the CO luminosities of the clumps range from 3 to 285\,K\,km\,s$^{-1}$\,pc$^{2}$, with an average of 
33\,K\,km\,s$^{-1}$\,pc$^{2}$ (see Table\,\ref{table:statistics of physical parameters}). 

\begin{figure}[t]
\centering
\includegraphics[width=0.5\textwidth]{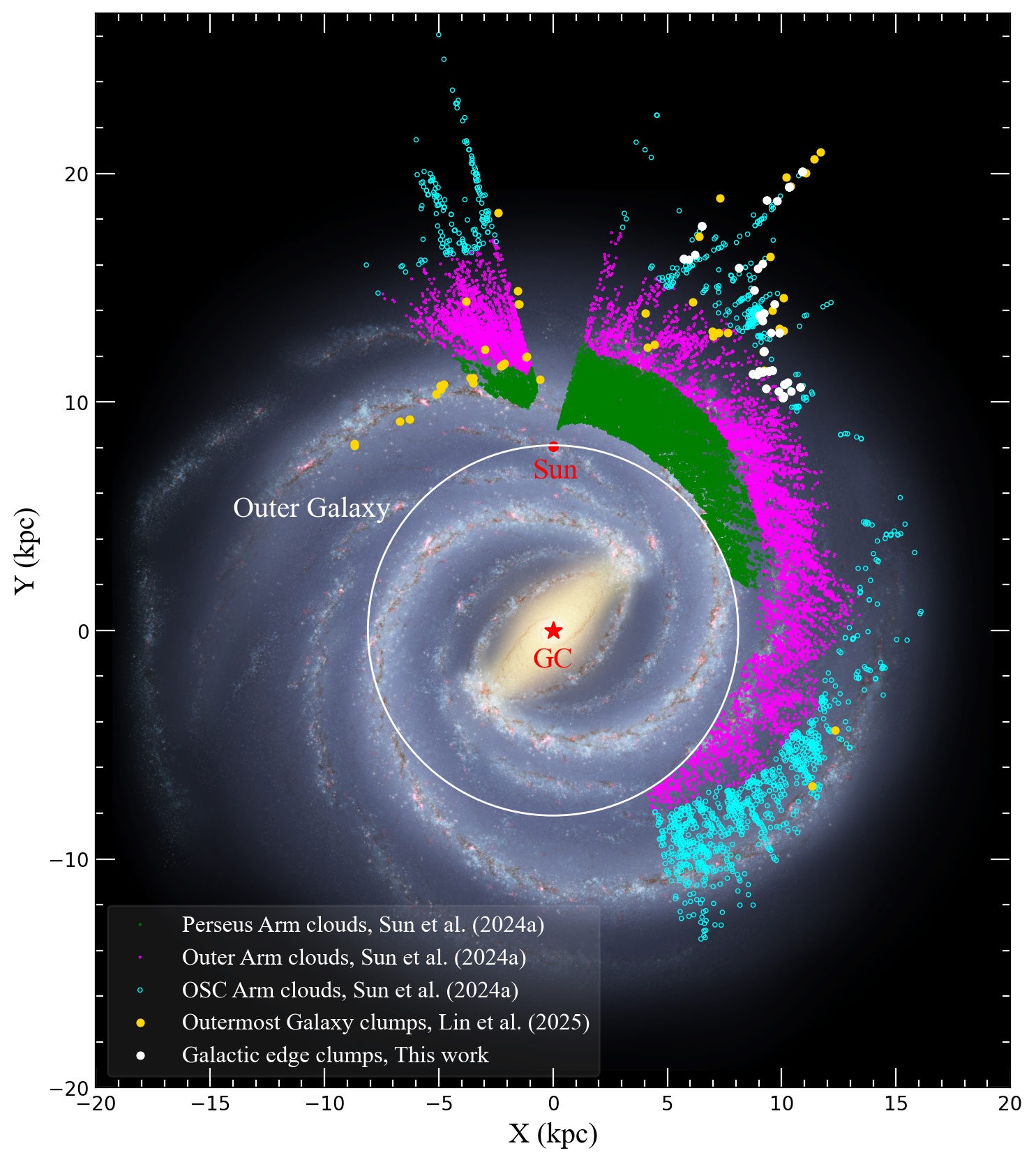} 
\caption{Locations of the molecular clouds discussed in Sect.\,\ref{sect:discussion}. 
The Perseus arm (\emph{green}), Outer arm (\emph{magenta}), OSC arm (\emph{cyan}) clouds are identified by 
the MWISP project with a beam size ($\theta_{\rm beam}$) of $\sim$\,50$^{\prime\prime}$ \citep{Sun2024b}. 
The gold and white solid circles denote Outermost Galaxy clumps ($\theta_{\rm beam}$\,$\sim$\,24$^{\prime\prime}$; \citealt{Lin2025}) 
and Galactic edge clumps ($\theta_{\rm beam}$\,$\sim$\,11$^{\prime\prime}$; \emph{this work}), respectively. 
The outer Galaxy refers to the region of the Galactic disk beyond the Solar circle (\emph{white circle}), 
including the Perseus arm, Outer arm, OSC arm, Outermost Galaxy, and Galactic edge. 
The background image is an artist's conception of the Milky Way, credited to R.Hurt: NASA/JPL-Caltech/SSC. 
}
\label{fig:target}
\end{figure}

\begin{figure*}[t]
\centering
\includegraphics[width=0.95\textwidth]{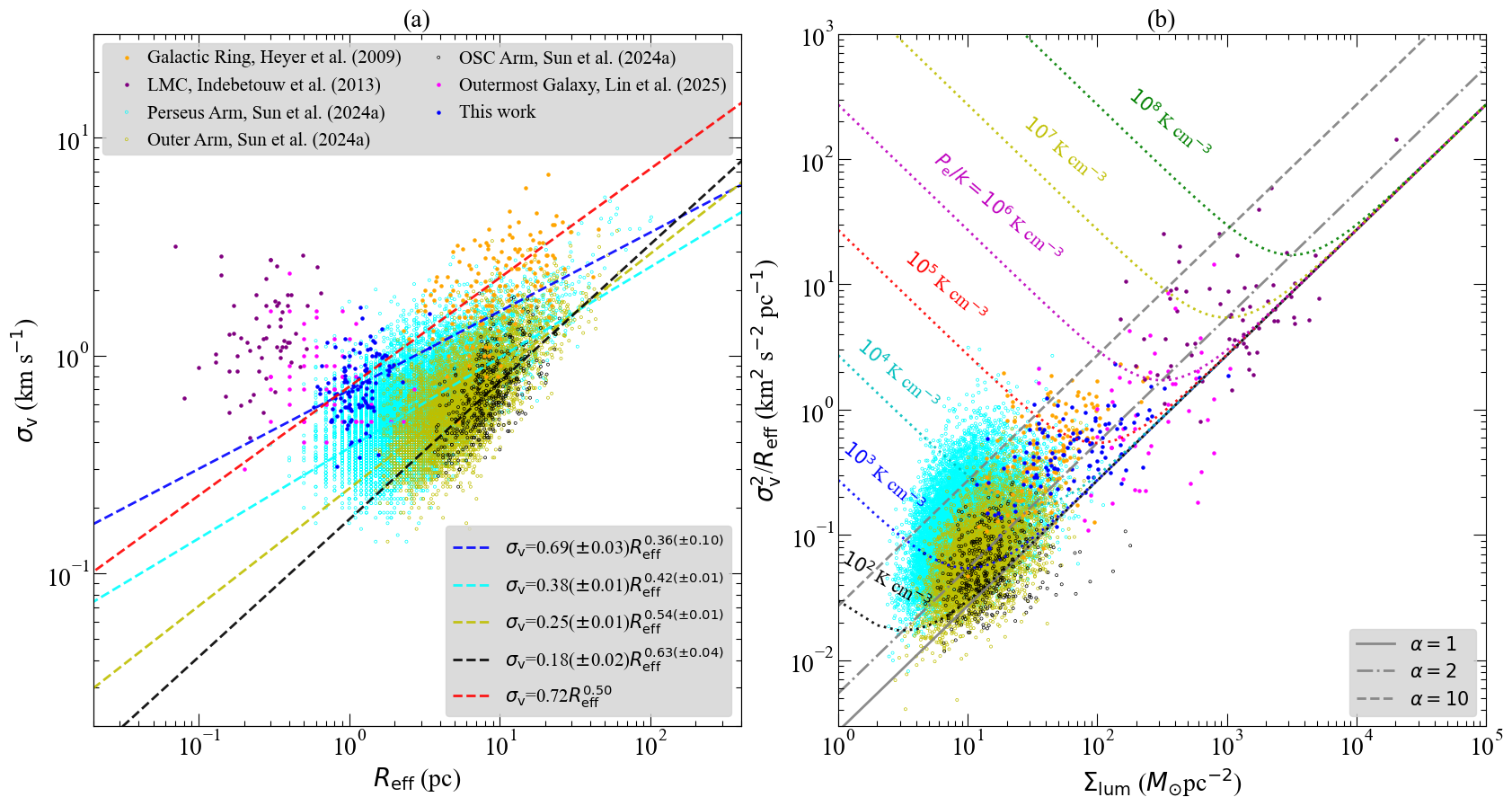}
\caption{(a) The $\sigma_v$--$R$ relations for molecular clouds located in the Galactic ring (\emph{orange}), 
Outermost Galaxy (\emph{magenta}), Perseus arm (\emph{cyan}), Outer arm (\emph{yellow}), OSC arm (\emph{black}), 
Galactic edge (\emph{blue}), and LMC (\emph{purple}). The blue, cyan, yellow, and black dashed lines represent the fits obtained for the Galactic edge clumps, Perseus arm clouds, Outer arm clouds, and OSC arm clouds, 
respectively. The dashed red line indicates the scaling relation for GMCs in the Milky Way
($R_{\rm g}$\,$\sim$\,0--10\,kpc; \citealt{Solomon1987}). (b) Scaling coefficient as a function of the mass surface density. 
The linear solid line represents virialized clouds in the absence of pressure ($\alpha_{\rm vir}$\,=\,1).
The linear dot-dashed and dashed lines correspond to $\alpha_{\rm vir}$ values of 2 and 10, respectively.
The dotted coloured curves indicate gravitational virial conditions for clouds confined by external pressures
($P_{\rm e}/k$) from $10^2$ to $10^8$\,K\,cm$^{-3}$ (for details, see Sect.\,\ref{sect:Stability of the clumps}). 
} 
\label{fig:size-velocity dispersion} 
\end{figure*}

Molecular gas masses are frequently estimated using CO\,(1--0) emission through the CO-to-H$_2$ conversion factor $X_{\rm CO}$. 
Clump masses are estimated from CO luminosity as, 
\begin{equation}
M_{\rm lum} = \alpha_{\rm CO} L_{\rm CO} \,,
\label{eq:luminous mass}
\end{equation}
where the locally determined mass-to-light ratio $\alpha_{\rm CO}$ is 4.3\,$\pm$\,1.3\,M$_{\odot}$\,(K\,km\,s$^{-1}$\,pc$^2$)$^{-1}$, 
corresponding to $X_{\rm CO}$\,=\,(2.0\,$\pm$\,0.6)\,$\times$\,$10^{20}$\,cm$^{-2}$\,(K\,km\,s$^{-1}$)$^{-1}$ \citep{Bolatto2013}. 
However, the $X_{\rm CO}$ factor is significantly influenced by metallicity. 
In this study, we have employed the relationship $X_{\rm CO}$\,$\propto$\,$Z^{-1.0}$ \citep{Arimoto1996,Bolatto2013,Lin2025}, 
where $Z$ signifies metallicity (${\rm log(}Z{\rm )}=-0.056R_{\rm g}-1.176$; for details, see \citealt{Giannetti2017}). 
Over the Galactocentric distance range spanned by our sample, this corresponds to metallicities of $Z$\,$\simeq $\,0.003--0.011, 
and hence to $X_{\rm CO}$ values of (4.3\,$\pm$\,0.7)--(13.4\,$\pm$\,0.8)\,$\times$\,10$^{20}$\,cm$^{-2}$\,(K\,km\,s$^{-1}$)$^{-1}$. 
The $X_{\rm CO}$ factor is also dependent on the clump radius \citep{Rubio1993}. 
This study does not account for the effect of clump size on the $X_{\rm CO}$ factor, given that the effective radii of 
our Galactic edge clumps are predominantly close to 1\,pc. 
This implies that our Galactic edge clumps may be spatially resolved. 
Using this approach, the derived clump masses span from 34 to 8250\,M$_\odot$, with a median value of 216\,$\pm$\,40\,M$_\odot$ 
(see Table\,\ref{table:statistics of physical parameters}). 
Most clump masses fall in the range of 80 to 500\,M$_\odot$ (see Fig.\,\ref{fig:pra_hist}). 

Using the CO-to-H$_2$ conversion factor, the average H$_2$ column densities $N({\rm H_2})$ are derived from the 
CO luminosities as 
\begin{equation}
N({\rm H_2}) = X_{\rm CO} L_{\rm CO} / (\pi R_{\rm eff}^2) \,. 
\label{eq:column density}
\end{equation}
The volume densities $n({\rm H_2})$ of the clumps are derived as follows: 
\begin{equation}
n({\rm H_2}) = 3 M_{\rm lum} / (4 \pi \mu m_{\rm H} R_{\rm eff}^3) \,, 
\label{eq:volume densities}
\end{equation}
where $m_{\rm H}$ denotes the mass of a hydrogen atom. 
$\mu$ indicates the mean molecular weight per H$_2$ in the interstellar medium. 
This value is typically 2.72, incorporating the mass of hydrogen, helium, and carbon monoxide isotopologues \citep{Li2015}. 
The $\mu$ values get slightly smaller at lower metallicities, yet the difference is marginal and can be considered negligible. 
The mean and median H$_2$ column densities of the CO clumps are 4.3\,$\times$\,10$^{21}$ and 2.9\,$\times$\,10$^{21}$\,cm$^{-2}$, 
respectively (see Fig.\,\ref{fig:pra_hist} and Table\,\ref{table:statistics of physical parameters}). 
Additionally, the volume density is mainly concentrated between 10$^{2}$ and 10$^{4}$\,cm$^{-3}$ with an average of $\sim$\,10$^{3}$\,cm$^{-3}$. 

Under the assumptions of an isothermal, spherical, and uniform system, and neglecting rotation, magnetic fields, 
and external pressure, the virial equilibrium condition can be written as \citep{McKee2007,Wong2022}: 
\begin{equation}
2T+W=2(\frac{3}{2}M_{\rm vir}\sigma_{\rm v}^2)-\frac{3}{5}\frac{{\rm G}M_{\rm vir}^2}{R_{\rm eff}}=0 \,,
\label{eq:virial balance}
\end{equation}
where $T$ and $W$ denote kinetic and potential energy, respectively. 
The virial mass of a clump is then given by  
\begin{equation}
M_{\rm vir} = 5 \sigma_{\rm v}^2 R_{\rm eff}/{\rm G} \,. 
\label{eq:virial mass}
\end{equation}
The surface density was calculated using the mass and radius of each individual clump as follows
\begin{equation}
\Sigma = M / (\pi R_{\rm eff}^2) \,. 
\label{eq:surface densities}
\end{equation}
Here, $\Sigma_{\rm lum}$ and $\Sigma_{\rm vir}$ are defined as the surface densities corresponding to 
$M_{\rm lum}$ and $M_{\rm vir}$, respectively. The surface densities ($\Sigma_{\rm lum}$) span from 12 
to 1025\,M$_{\odot}$\,pc$^{-2}$, with a mean of 93\,$\pm$\,19 and a median of 63\,$\pm$\,13\,M$_{\odot}$\,pc$^{-2}$ 
(see Table\,\ref{table:statistics of physical parameters}). 

The virial parameter serves as a crucial indicator of the dynamical state of molecular clumps, 
quantifying the ratio of internal kinetic energy to gravitational energy. 
In this study, we adhere to the definition provided by \citet{Bertoldi1992}, 
\begin{equation}
\alpha_{\rm vir} = M_{\rm vir}/M_{\rm lum} \,. 
\label{eq:virial parameter}
\end{equation}
The derived virial parameters of CO clumps range from 0.6 to 15.3, with a mean of 3.4\,$\pm$\,0.8 
and a median of 2.8\,$\pm$\,0.6 (see Table\,\ref{table:statistics of physical parameters}).
The physical parameters of the 112 identified CO clumps are summarized in Table\,\ref{table:parameters}. 
The distributions of the physical parameters are shown in Fig.\,\ref{fig:pra_hist}. 
Statistical analyses are performed for all clumps, focusing on properties such as velocity dispersion,
size, luminosity, mass, and density (see Table\,\ref{table:statistics of physical parameters}). 

\begin{figure*}[t]
\centering
\includegraphics[width=0.95\textwidth]{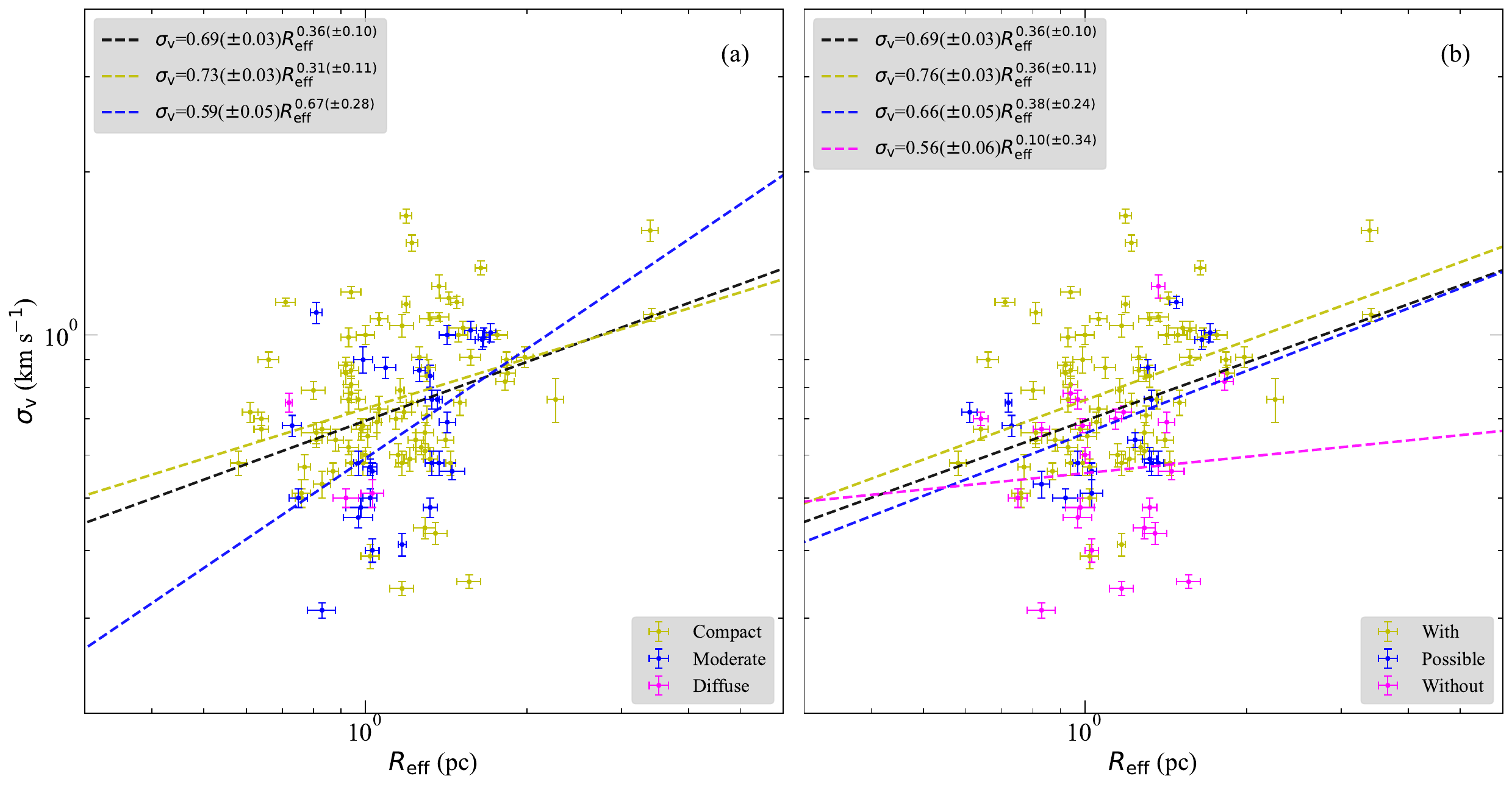}
\caption{(a) The $\sigma_v$--$R$ relation for the Galactic edge clumps identified in compact (\emph{yellow}), 
intermediate (\emph{blue}), and diffuse clouds (\emph{magenta}). The black, yellow, and blue dashed lines
represent the fits obtained for our Galactic edge clumps identified in all, compact, and intermediate clouds, 
respectively. Fitting of diffuse clouds is not performed due to only three values. 
(b) The $\sigma_v$--$R$ relation for the Galactic edge clumps, all of them (\emph{black}), clouds with star 
forming activity (\emph{yellow}), with possible star-forming activity (\emph{blue}), and without star-forming activity (\emph{magenta}). 
The classification for clouds is detailed in Figs.\,\ref{fig:map} and \ref{fig:map2} or Table\,C.1 of \citealt{Luo2025}.} 
\label{fig:size-linewidth-class} 
\end{figure*}

\subsection{Uncertainty estimation} 
\label{sect:Error propagation} 
Uncertainties in the above derived parameters were calculated through error propagation and were primarily determined by the uncertainties in kinematic distance, velocity dispersion, and CO luminosity. 
Uncertainties in the kinematic distance derived from the Galactic rotation curve propagate to the clump parameters and scaling relationship fits. 
Uncertainty in the physical size ($\sigma_{R_{\rm eff}}$) of clumps is proportional to the adopted heliocentric distance uncertainty ($\sigma_{D_{\rm s}}$), $\sigma_{R_{\rm eff}}$\,$\propto$\,$\sigma_{D_{\rm s}}$, assuming that the angular size uncertainty is negligible. 
Given a typical heliocentric distance uncertainty of $\sim$\,3--5\%, the physical size exhibits a comparable level of uncertainty (see Table\,\ref{table:statistics of physical parameters}). 
In addition, uncertainty in the Galactocentric radius propagates to the metallicity and thus to the adopted CO-to-H$_2$ conversion factor. 
In our analysis (see Sect.\,\ref{sect:Estimation of Physical Parameters}), this effect is incorporated into the first term of the luminous mass uncertainty $\sigma_{M_{\rm lum}}$ as 
\begin{equation}
\sigma_{M_{\rm lum}} = M_{\rm lum} \sqrt{(\frac{\sigma_{X_{\rm CO}}}{X_{\rm CO}})^2+(2\frac{\sigma_{D_{\rm s}}}{D_{\rm s}})^2+(\frac{\sigma_{R_{21}}}{R_{21}}})^2 \,. 
\label{eq:surface densities}
\end{equation}
Here, $\sigma_{X_{\rm CO}}$ accounts for the propagation of uncertainty in metallicity resulting from the Galactocentric radius uncertainty ($\sigma_{R_{\rm g}}$), where $\sigma_{X_{\rm CO}}$\,=\,0.129$X_{\rm CO}\sigma_{R_{\rm g}}$. 
The $\sigma_{D_{\rm s}}$ and $\sigma_{R_{21}}$ represent the uncertainties in heliocentric distance and the CO\,$J$=2--1/1--0 line ratio, respectively. 
These uncertainties are propagated to all derived quantities (e.g., $L_{\rm CO}$, $M_{\rm lum}$, $\Sigma_{\rm lum}$, $\alpha_{\rm vir}$), encompassed within the reported errors and subsequent scaling relation fits (see Sect.\,\ref{sect:discussion}). 

\section{Discussion}
\label{sect:discussion}
\subsection{Clump parameters as a function of distance}
\label{sect:Clump parameters as a function of distance} 
Previous observations indicate that molecular clouds in the outer Galaxy, observed in CO, are generally smaller, 
less massive, and exhibit lower densities and narrower linewidths compared to those in the inner Galaxy 
(e.g., \citealt{Brand1995,May1997,Heyer2015,Sun2024a,Lin2025}). 
As mentioned in Sect.\,\ref{sect:Introduction}, the differences in physical properties between the inner 
and outer Galaxy can be attributed to variations in the surrounding environment. 
In the outer Galaxy, CO-dark H$_2$ gas may contribute more significantly than locally to the total molecular mass, 
potentially introducing systematic uncertainties in mass-related estimates (see Sect.\,\ref{sect:conversion factor}). 
A radial gradient has been reported in cloud velocity dispersion ($R_{\rm g}$\,=\,0--24\,kpc; \citealt{Sun2024a,Lin2025}). 
Observed surface densities also appear to show a radial decrease, diminishing from 1800\,M$_{\odot}$\,pc$^{-2}$ 
near the Galactic center to 200\,M$_{\odot}$\,pc$^{-2}$ in the molecular ring ($R_{\rm g}$\,$\sim$\,3--5\,kpc) 
and further down to about 30\,M$_{\odot}$\,pc$^{-2}$ in the outer Galaxy 
($\theta_{\rm beam}$\,$\sim$\,45$^{\prime\prime}$; see Fig.\,8 in \citealp{Heyer2015}). 
In contrast, \citealt{Urquhart2024} report no discernible difference in the surface density of clumps between
the inner and outer Galaxy ($\theta_{\rm beam}$\,$\sim$\,30$^{\prime\prime}$). 
The distributions of clump mass and luminosity seem to show a minimum near the Solar circle ($\sim$\,8\,kpc), but this is attributable to the particularly high sensitivity and resolution accessible for sources near the Sun \citep{Urquhart2024}. 
Beyond $\sim$\,14\,kpc from the Galactic center, the CO luminosity no longer appears to exhibit a clear 
declining trend  due to the increasing detection threshold with heliocentric distance \citep{Heyer2015}. 

Compared to previous studies, our data provide a substantially higher angular resolution 
($\theta_{\rm beam}$\,$\sim$\,11$^{\prime\prime}$), which is particularly relevant for size- and 
density-related parameters. 
Taking into account the variation of the CO-to-H$_2$ conversion factor with Galactocentric distance, 
as outlined in Sect.\,\ref{sect:Estimation of Physical Parameters}, the ranges of size, mass, 
and volume density of our Galactic edge clumps are found to be 0.6--3.4\,pc, 34--8250\,M$_\odot$, 
and 100--16100\,cm$^{-3}$, respectively
(see Tables\,\ref{table:statistics of physical parameters} and \ref{table:parameters}). 
Over the Galactocentric distance range of 14--23\,kpc, we find no compelling evidence for systematic 
radial trends in clump size, mass, volume density, velocity dispersion, luminosity, or surface density
(see Fig.\,\ref{fig:pra_dis}). No clear dependence of these properties on heliocentric distance is 
observed over $D_{\rm s}$\,$\sim$\,9--16\,kpc. We observe marginal increases in clump size and mass at 
the largest distances ($R_{\rm g}$\,$\sim$\,22--23\,kpc, corresponding to $D_{\rm s}$\,$\sim$\,15--16\,kpc; see Fig.\,\ref{fig:pra_dis}), 
while the volume density remains approximately constant. Notably, the mean clump mass in the 
outermost Galactocentric bins ($\sim$\,1863\,M$_\odot$; see Fig.\,\ref{fig:pra_dis}\,(c)) is nearly 
an order of magnitude higher than the median mass of the full sample 
(216\,M$_\odot$; see Table\,\ref{table:statistics of physical parameters}).
This apparent enhancement is most likely driven by distance-related selection effects. 
These elevated values are associated with clumps identified in compact star-forming molecular clouds (e.g., G137.759$-$0.983 and G137.775$-$1.067; see Fig.\,\ref{fig:map2} and \citealt{Luo2025}). 

The angular resolution of our observations is significantly higher than that of most previous outer Galaxy studies, 
which may introduce systematic differences when comparing derived properties. 
The CO-to-H$_2$ conversion factor depends not only on the metallicity but also on the beam size
(see Sect.\,\ref{sect:conversion factor}). 
Higher resolution tends to yield lower $X_{\rm CO}$ values, owing to the discernible denser substructure
and stronger peak brightness \citep{Rubio1993}. 
In addition, peak column densities and surface densities are sensitive to beam dilution effects. 
Nevertheless, integrated properties (e.g., luminosity, mass) of the clumps exhibit relative robustness, 
although minor adjustments may be necessary for very small clouds. 
Global properties such as velocity dispersion, virial parameter, and scaling relations are less sensitive 
to beam size as long as the structure is resolved 
(see Sects.\,\ref{sect:Size-line-width Relation}, \ref{sect:Mass-size Relation}, and \ref{sect:Stability of the clumps}). 

\subsection{Velocity dispersion-size relation}
\label{sect:Size-line-width Relation}
As mentioned in Sect.\,\ref{sect:Introduction}, the first Larson relation demonstrates a power-law 
form between velocity dispersion and size, represented as $\sigma_{\rm v}$\,=\,$v_0 R_{\rm eff}^{\,m}$. 
The $\sigma_v$--$R$ relation is considered as a result of multi-scale turbulent motions within the interstellar medium. 
Large-scale turbulence promotes the fragmentation of molecular clouds, subsequently leading to the formation of clumps. 
Conversely, small-scale turbulence may suppress star formation by sustaining gravitational collapse \citep{Larson1981}. 
This relation gains increasing credibility due to the utilization of multi-tracers and large cloud samples. 
The $\sigma_v$--$R$ relation of molecular clouds has been studied in various tracers, including CO, H$_2$CO,
HCO$^{+}$, HCN, N$_2$H$^{+}$, CS, and NH$_3$ 
(e.g., \citealt{Larson1981,Fuller1992,Goodman1998,Shetty2012,Kauffmann2017,Gong2023,Luo2024b}). 
Furthermore, it has been studied through observations on various scales, including cloud scale
(e.g., \citealt{Larson1981,Solomon1987}), clump scale (e.g., \citealt{Traficante2018,Luo2024b}), 
and core scale (e.g., \citealt{Fuller1992,Caselli1995,Barnes2021,Li2023}). 
The $\sigma_v$--$R$ relation in low metallicity galaxies has been found to be consistent with that 
of the Milky Way, where the coefficient $v_0$ is highest for the Milky Way and the index $m$ is
highest for dwarf galaxies \citep{Rubio2015}. 
Typical values found in previous studies are $v_0$\,$\sim$\,0.4--3.0 and $m$\,$\sim$\,0.3--0.6, 
depending on the tracer, spatial scale, and environment. 

\begin{figure}[t]
\centering
\includegraphics[width=0.5\textwidth]{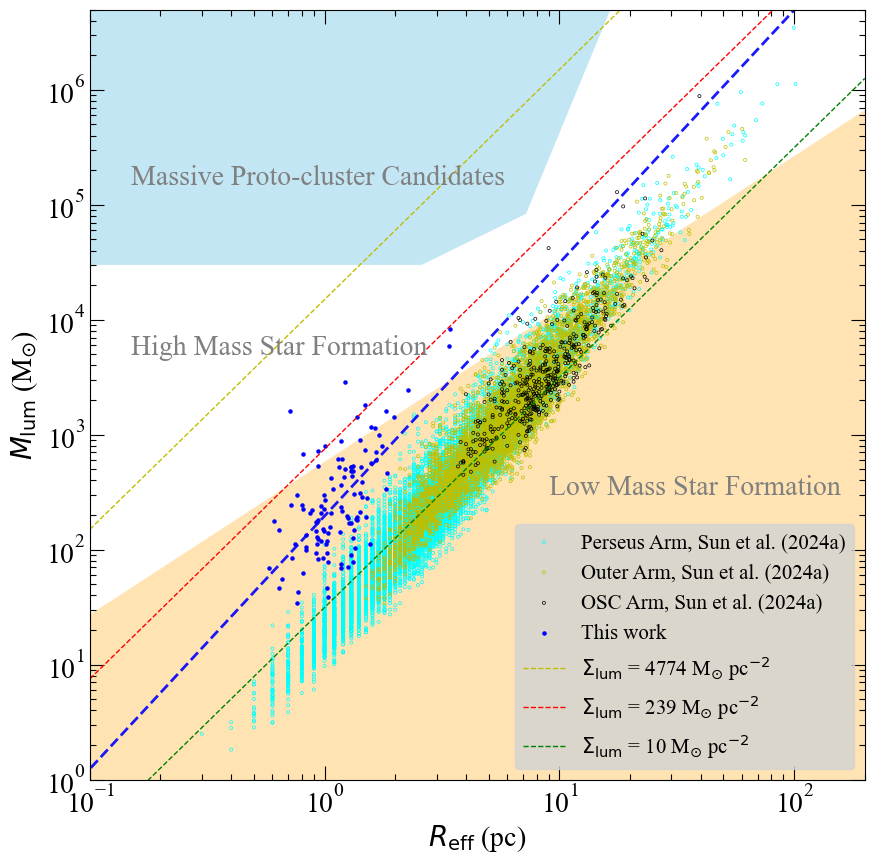}
\caption{The $M$--$R$ relation for molecular clouds. Results are obtained from Perseus arm clouds
(\emph{cyan}), Outer arm clouds (\emph{yellow}), OSC arm clouds (\emph{black}), 
and Galactic edge clumps (\emph{blue}). The orange-shaded region represents the parameter space
associated with low mass star formation, 
where $M_{\rm lum}$\,$\leq$\,580M$_{\odot}$($R_{\rm eff}$/pc)$^{1.33}$ \citep{Kauffmann2010c}. 
The cyan-shaded region represents the parameter space associated with massive proto-cluster candidates \citep{Bressert2012}. 
The blue dashed line represents the fitted result for this work ($M_{\rm lum}$\,=\,196$R_{\rm eff}^{\,2.18}$). 
The green, red, and yellow dashed lines represent surface densities of 10, 239, and 4774\,M$_{\odot}$\,pc$^{-2}$, respectively. }  
\label{fig:Reff-Mlum} 
\end{figure}

\begin{figure*}[t]
\centering
\includegraphics[width=0.95\textwidth]{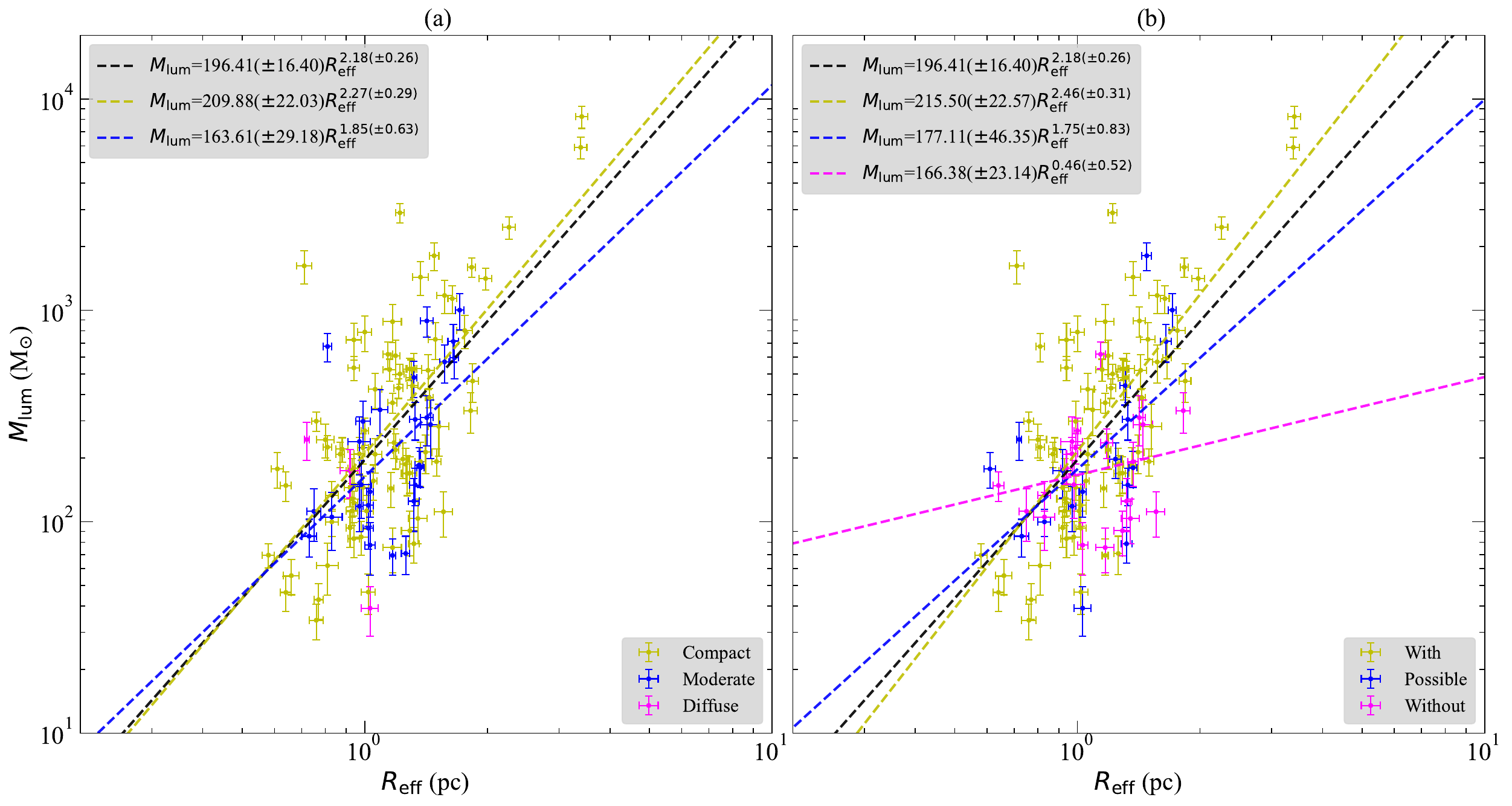}
\caption{The $M$--$R$ relation for our Galactic edge clumps. 
The detailed descriptions are the same as in Fig.\,\ref{fig:size-linewidth-class}. 
} 
\label{fig:size-mass-class} 
\end{figure*}

On scales 0.1--100\,pc, the physical parameters of molecular clouds consistently adhere to the $\sigma_v$--$R$ 
relation across various environments (see Fig.\,\ref{fig:size-velocity dispersion}). 
In this work, we focus on the $\sigma_v$--$R$ relation of CO\,(2--1) clumps within the Galactic edge clouds. 
The $\sigma_v$--$R$ relation is fitted by the equation $\sigma_{\rm v}$\,=\,0.69($\pm$0.03)$R_{\rm eff}^{0.36(\pm0.10)}$. 
The Pearson's coefficient of 0.34 indicates a weak correlation, and the $p$-value less than $10^{-3}$ indicates high significance (see Table\,\ref{table:Fit results for the scaling relations}). 
The index $m$ of 0.36 in our Galactic edge clumps is closer to Larson's value of 0.38 \citep{Larson1981}, compared to the subsequently determined value of 0.50 \citep{Solomon1987}. 
Outer-Galaxy molecular clouds (15$^\circ$\,<\,$l$\,<\,165$^\circ$ and targets at 
195$^\circ$\,<\,$l$\,<\,230$^\circ$, -5$^\circ$\,<\,$b$\,<\,5$^\circ$, 8\,<\,$R_{\rm g}$\,<\,26\,kpc) 
identified by the MWISP survey in the Perseus, Outer, and Outer
Scutum-Centaurus (OSC) arms exhibit $\sigma_v$--$R$ scalings with fitted indices $m$ close to 0.5 
(see Figs.\,\ref{fig:target} and \ref{fig:size-velocity dispersion}; \citealt{Sun2024a,Sun2024b}). 
However, for our Galactic edge clumps the fitted index $m$ of 0.36 is lower but the coefficient $v_0$ of 
0.69 is larger than in previous studies \citep{Sun2024a}. 
The velocity dispersion within our Galactic edge clumps is analogous to that of the Outermost Galaxy 
clumps of the same size, as derived from $^{13}$CO ($\theta_{\rm beam}$\,$\sim$\,24$^{\prime\prime}$; \citealt{Lin2025}). 
Molecular clouds found in the outer Galaxy disk are characterized by low velocity dispersion. 
The Galactic ring clouds, as observed from $^{13}$CO\,(1--0) data, also adhere to the $\sigma_v$--$R$ 
relation \citep{Heyer2009}, despite their systematically larger velocity dispersion at a given size 
compared to those in the outer Galaxy (see Fig.\,\ref{fig:size-velocity dispersion}). 

Star formation activity appears to escalate as the velocity dispersion increases from small to 
large at a specified size within the the Large Magellanic Cloud (LMC). 
Previous observations suggest that the velocity dispersions in 30\,Dor are much larger than those in N159, which may be due to the 
particularly vigorous star-formation activity in 30\,Dor compared to N159 \citep{Indebetouw2013,Tang2017a,Tang2021,Nayak2018}. 
One could infer that a greater velocity dispersion at a fixed size might indicate intensified turbulence. 
High velocity dispersion is a result of the energetic feedback from recent star formation,
or the collapse into stellar objects due to gravity-driven motions 
(e.g., \citealt{Wong2019,Tang2017a,Tang2017b,Tang2018a,Tang2018b,Tang2021}). 
However, in the Galactic center region star formation is inhibited due to too high turbulence and 
sheer triggered by differential rotation \citep{Lu2019}. 
This may indicate the presence of a turbulence threshold above which star formation is suppressed. 

The fitted index $m$ is 0.31\,$\pm$\,0.11 for compact clouds and 0.67\,$\pm$\,0.28 for intermediate clouds.
The fitted results indicate a weak but very significant correlation for compact clouds and a moderate and statistically significant correlation for intermediate clouds (see Fig.\,\ref{fig:size-linewidth-class} and Table\,\ref{table:Fit results for the scaling relations}). 
The compact clouds likely exhibit a lower index $m$ compared to intermediate clouds, suggesting a slower 
increase in velocity dispersion with scale. 
In contrast, for clouds with intermediate structure, larger scales lead to a more rapid increase in 
velocity dispersion, resulting in a steeper slope of the $\sigma_v$--$R$ relation. 
Additionally, the fitted coefficient $v_0$ is 0.73\,$\pm$\,0.03 for compact clouds and 0.59\,$\pm$\,0.05 
for intermediate clouds. 
Among our 72 Galactic edge clouds, 26 were identified to host star formation activity, while 13 might be
associated with star formation activity \citep{Sun2015b,Luo2025}. 
The fitted index $m$ for clouds with, possible, and without star-formation activity are 
0.36\,$\pm$\,0.11, 0.38\,$\pm$\,0.24, and 0.10\,$\pm$\,0.34, respectively. 
The corresponding Pearson's coefficients are 0.39, 0.44, and 0.06, with the $p$-values of less than $10^{-3}$, 0.075, and 0.782, respectively (see Fig.\,\ref{fig:size-linewidth-class} and Table\,\ref{table:Fit results for the scaling relations}). 
This indicates that the $\sigma_v$--$R$ correlation is much weaker and insignificant in molecular clouds without star formation. 
The index $m$ of clouds with possible star formation activity appears to be equivalent to that
with certain star formation activity, but with larger uncertainties. 
The fitted coefficients $v_0$ derived from our Galactic edge clumps within molecular clouds, 
which are associated with star-formation activity (0.76\,$\pm$\,0.03), exhibit values that 
are consistently higher than those obtained in scenarios with both possible (0.66\,$\pm$\,0.05)
and without star-formation activity (0.56\,$\pm$\,0.06) (see Fig.\,\ref{fig:size-linewidth-class}). 
This suggests that the $v_0$ value could be elevated in scenarios of vigorous turbulence-induced 
star formation, not only in the LMC but also in the outer Galaxy. 
Despite substantial uncertainties, the apparent difference in $m$ and $v_0$ could indicate distinct degrees of internal dynamics in molecular clouds with different overall properties, including compactness levels with/without star formation activity. 

\subsection{Mass-size relation}
\label{sect:Mass-size Relation}
A relation between mass and size is referred to as the third Larson relation. 
Molecular clouds maintain an almost constant surface density, denoted as $n$\,$\propto$\,$L^{\rm -1.1}$ \citep{Larson1981}. 
This can be converted to $M_{\rm lum}$\,$\propto$\,$R_{\rm eff}^{\rm 1.9}$. 
A similar power-law relation has been observed in molecular clouds both in the Milky Way 
(e.g., \citealt{Miville-Deschenes2017,Traficante2018,Lada2020,Ma2021,Xing2022,Sun2024a}) and 
nearby galaxies (e.g., \citealt{Mookerjea2004,Reid2006,Faesi2018,Lada2024}). 
The multi-scale $M$--$R$ relation, ranging from cores to clouds (0.05--10\,pc), 
has been thoroughly investigated by \citet{Kauffmann2010a,Kauffmann2010b}. 
In the case of the $M_{\rm lum}$\,$\propto$\,$R_{\rm eff}^{\,q}$ relation, the mean column density 
decreases in clouds with larger radii, when $q$\,<\,2. 
Conversely, there is an increase in column density, when the cloud size increases for $q$\,>\,2. 
The $q$-value was reported to be larger than 2 in the infrared dark clouds on scales of 0.01--0.1\,pc \citep{Ragan2009}. 
The $q$ value is less than 2 in the barred galaxy NGC\,1300 on a 100\,pc scale \citep{Maeda2020}. 
This suggests that the mean surface density may decrease with increasing cloud radius on large scales, 
while it increases as the cloud radius increases on small scales. 
The clumps are primarily supported by turbulence for $q$\,=\,2, while the clump's volume density
is uniform, i.e., size independent, for $q$\,=\,3 \citep{Larson1981}. 

\begin{figure*}[t]
\centering
\includegraphics[width=0.95\textwidth]{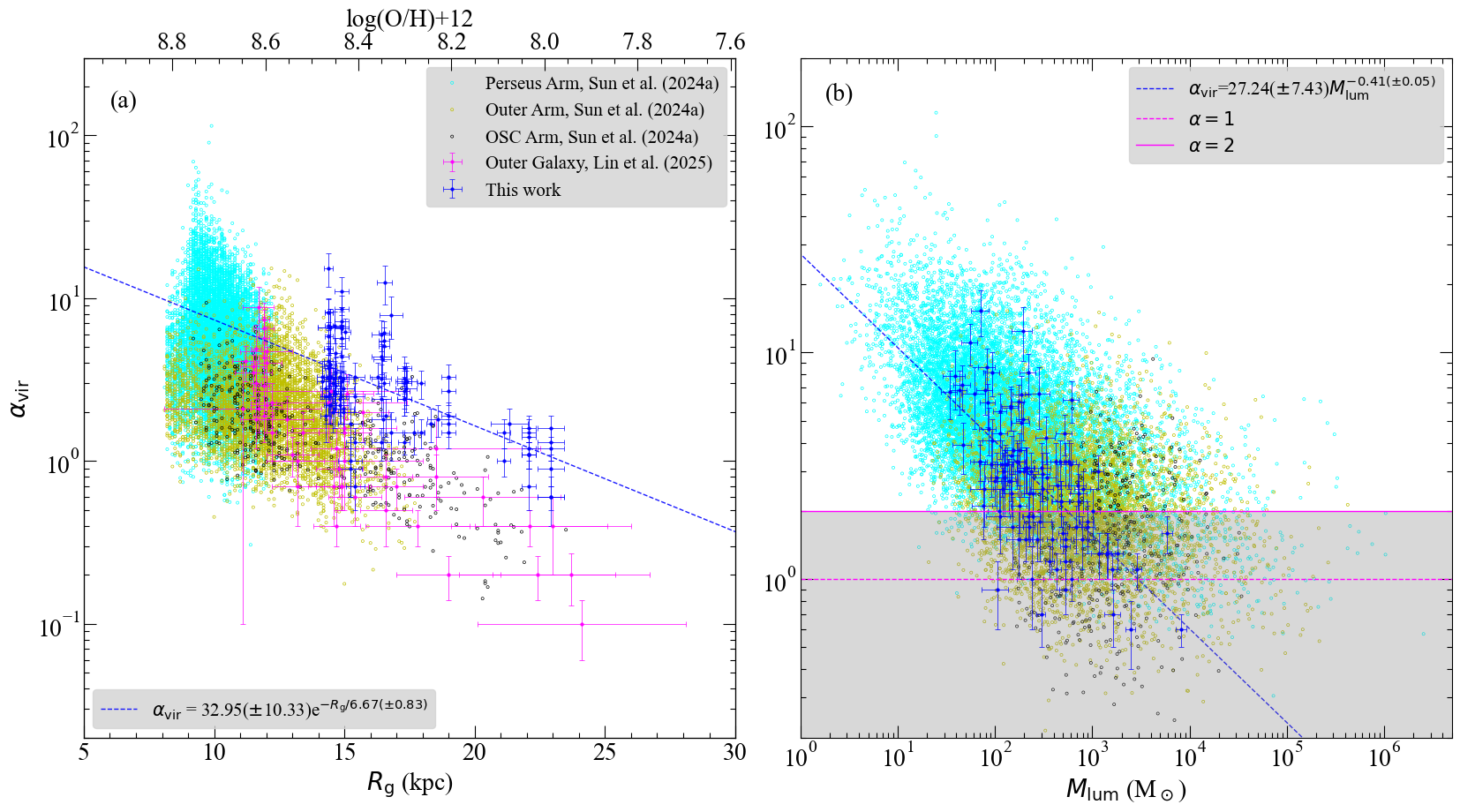}
\caption{(a) Variation of cloud virial parameter with Galactocentric distance. 
Results are obtained from Perseus arm clouds (\emph{cyan}), Outer arm clouds (\emph{yellow}), OSC arm clouds (\emph{black}), 
Outermost Galaxy clumps (\emph{magenta}), and our Galactic edge clumps (\emph{blue}). 
The blue dashed line represents the fitted result for our Galactic edge clumps ($\alpha_{\rm vir}$\,=\,33.0e$^{-R_{\rm g}/6.7}$). 
(b) Variation of cloud virial parameter with CO luminous mass. 
The horizontal magenta dashed and solid lines denote $\alpha_{\rm vir}$\,=\,1 and 2, respectively. 
The blue dashed line represents the fitted result for our Galactic edge clumps 
($\alpha_{\rm vir}$\,=\,27.24$M_{\rm lum}^{\,-0.41}$). } 
\label{fig:Rg-alpha} 
\end{figure*}

Within our Galactic edge clumps (see Fig.\,\ref{fig:Reff-Mlum} and Table\,\ref{table:Fit results for the scaling relations}), the $M$--$R$ relation is fitted to $M_{\rm lum}$\,=\,196($\pm$17)$R_{\rm eff}^{\,2.18\,(\pm0.26)}$. 
The Pearson's coefficient is with 0.63 much higher than in the case of the $\sigma_v$--$R$ 
correlation, indicating a strong positive connection. 
And the $p$-value less than $10^{-3}$ indicates high significance. 
The value of $q$ close to 2 implies that the clumps are turbulence-supported and maintain a roughly constant average column density, consistent with the weak dependence of column density on $R_g$ observed in Fig.\,\ref{fig:pra_dis}. 
As the scale increases, clumps maintain a stable column density, which may be regulated by external pressure or turbulent driving forces. 
The fitted index $q$\,=\,2.18 of our Galactic edge clumps is slightly lower than the value of 2.63 
reported by \citet{Sun2024a} for outer-Galaxy molecular clouds at $R_{\rm g} \ge 12.5$\,kpc, 
likely reflecting the difference between cloud- and clump-scale measurements. 
The coefficient of the $M$--$R$ relation shows considerable variation between the inner and outer Galaxy, 
corresponding to different surface densities \citep{Lada2020}. 
The mean surface density of our Galactic edge clumps 
($\sim$\,92.9\,M$_{\odot}$\,pc$^{-2}$; see Table\,\ref{table:statistics of physical parameters}) 
is larger than previously reported values for the outer Galaxy clouds
($\sim$30\,M$_{\odot}$\,pc$^{-2}$ at $\theta_{\rm beam}$\,$\sim$\,45$^{\prime\prime}$; \citealt{Heyer2001}). 
Furthermore, the average surface density within molecular clouds situated in the Perseus arm, 
Outer arm, and OSC arm is measured at 8.1, 8.3, and 7.5\,M$_{\odot}$\,pc$^{-2}$, respectively
($\theta_{\rm beam}$\,$\sim$\,50$^{\prime\prime}$; \citealt{Sun2024a}). 
These findings suggest that the surface density of molecular clouds in the outer Galaxy is
lower than that in the inner Galaxy. Nevertheless, when considering the different fractions of 
CO-dark gas, the discrepancies are not as large as those derived from a constant $X_{\rm CO}$ value. 
The fraction of CO-dark H$_2$ gas ($f_{\rm dark}$) increases from $\sim$\,20\% at $R_{\rm g}$\,=\,4\,kpc to $\sim$\,80\% at $R_{\rm g}$\,=\,10\,kpc \citep{Pineda2013}. 
For the total surface density ($\Sigma_{\rm total}$) of molecular clouds, the CO-derived surface density ($\Sigma_{\rm CO}$) could be corrected by $\Sigma_{\rm total}$\,=\,$\Sigma_{\rm CO}$/(1-$f_{\rm dark}$), corresponding to correction factors of $\sim$\,1.25 in the inner Galaxy and up to $\sim$\,5 in the outer Galaxy. 
In the outer Galaxy clouds, massive star formation and massive proto-cluster candidates are notably scarce. 
This could imply a tendency toward a bottom-heavy stellar initial mass function (IMF) in Galactic edge clumps. 
While similar trends have been suggested for low-metallicity galaxies such as the Small Magellanic Cloud (SMC), the relatively high $^{18}$O/$^{17}$O ratios in the outer Galaxy indicate distinct chemical conditions. 
Therefore, the IMF behavior in the outer Galaxy may not directly follow that observed in the SMC (e.g., \citealt{Ruffle2007,Zou2023,Gong2025}). 
This may in part be influenced by large mass contributions of CO-dark gas. 
Another possibility is that the outer Galaxy, being relatively young, has had not yet sufficient time to agglomerate a high amount of $^{17}$O from stars of intermediate mass. 

The fitted index $q$ is 2.27\,$\pm$\,0.29 for compact clouds and 1.85\,$\pm$\,0.63 for intermediate 
clouds. 
The fitted results indicate a strong and highly significant correlation for compact clouds, as well as a moderate but very significant correlation for intermediate clouds (see Fig.\,\ref{fig:size-mass-class} and Table\,\ref{table:Fit results for the scaling relations}). 
Similarly, the fitted index $q$ for clouds with, possible, and without star-formation activity are 2.46\,$\pm$\,0.31, 1.75\,$\pm$\,0.83, and 0.46\,$\pm$\,0.52, respectively. 
The Pearson's coefficient of 0.22 ($p$-value = 0.323) indicates that the $M$--$R$ relation is substantially weaker and statistically insignificant in molecular clouds devoid of star formation. 
Star forming and possibly star forming clouds indicate that $q$ is close to 2, although the values 
for possibly star forming clouds are subject to significant uncertainty. 
In contrast, the no star forming clouds may have a much lower $q$ value. 
In addition, the fitted coefficient of the $M$--$R$ relation is 210\,$\pm$\,23 for compact clouds 
and 164\,$\pm$\,30 for intermediate clouds (see Table\,\ref{table:Fit results for the scaling relations}). 
This suggests that compact clouds may contain more mass at a given size, and thus have higher surface 
densities than intermediate clouds. 
The fitted coefficient of the $M$--$R$ relation for clouds with, 
possible, and without star-formation activity are 216\,$\pm$\,23, 177\,$\pm$\,47, 166\,$\pm$\,24, respectively. 
This indicates that the gas surface density may increase during the star formation process, 
but the coefficient for possible star-formation activity clouds remains highly uncertain. 
As listed in Table\,\ref{table:Fit results for the scaling relations}, the coefficient of the $M$--$R$ relation within molecular clouds, 
situated in the Perseus arm (19.3\,$\pm$\,0.2), Outer arm (20.8\,$\pm$\,0.5), and OSC arm (30.9\,$\pm$\,4.3), is notably lower than that of our clumps (196\,$\pm$\,17). 
It should be noted that the typical spatial resolution of our Galactic edge clumps ($\sim$\,0.5--0.9\,pc) 
surpasses that of the Perseus arm ($\sim$\,1\,pc), the Outer arm ($\sim$\,3\,pc), and the OSC arm ($\sim$\,5\,pc) clouds. 
One would expect to observe higher density values in instances of greater resolution. 

\subsection{Virial stability} 
\label{sect:Stability of the clumps}
The equilibrium state of the clumps is typically characterized by the ratio of internal kinetic energy
to gravitational potential energy (i.e., virial parameter $\alpha_{\rm vir}$; see Eq.\,\ref{eq:virial parameter}), 
a concept known as the second Larson relation \citep{Larson1981}. 
The critical virial parameter ($\alpha_{\rm cri}$) is used to determine whether a molecular cloud 
can maintain equilibrium. In the ideal case of virial equilibrium, $\alpha_{\rm vir} \approx 1$, 
while values up to $\alpha_{\rm vir} \sim 2$ are often considered consistent with gravitationally 
bound structures when accounting for non-ideal effects such as turbulence and external perturbations \citep{Kauffmann2013,Rigby2019}. 
Approximately 80\% of cores situated within 70\,$\mu$m dark high-mass clumps in infrared dark clouds 
are gravitationally bound ($\alpha_{\rm vir}$\,<\,2; \citealt{Li2023}). 
However, the existence of gravitationally bound dense cores does not necessitate that the parent 
cloud be globally bound. The virial parameter of our Galactic edge clumps spans a wide range from 
0.6 to 15.3, with most values concentrated between 1 and 4 (see Fig.\,\ref{fig:pra_hist} and 
Table\,\ref{table:statistics of physical parameters}). The median virial parameter is 2.8\,$\pm$\,0.6, 
indicating that the majority of clumps possess kinetic energies exceeding their gravitational binding 
energies and are therefore not gravitationally bound. Only 28\% of the clumps exhibit $\alpha_{\rm vir}$ 
values between 1 and 2, while merely 5\% have $\alpha_{\rm vir}$\,<\,1 (see Table\,\ref{table:parameters}). 
Clumps with $\alpha_{\rm vir}$\,<\,2 are predominantly associated with compact clouds (74\%), 
compared to intermediate (20\%) and diffuse clouds (6\%), suggesting that clump stability is closely linked to cloud compactness. 
Although uncertainties in the
CO-to-H$_2$ conversion factor and the presence of CO-dark molecular gas in the
low-metallicity Galactic edge may affect the absolute values of $\alpha_{\rm vir}$
(see Sect.\,\ref{sect:conversion factor}), most clumps remain clearly supervirial even
under conservative assumptions. 

To assess the role of external pressure, we consider pressure-bounded virial equilibrium for an 
isothermal, self-gravitating spherical cloud embedded in a uniform external pressure $P_{\rm e}$ \citep{Field2011},
\begin{equation}
v_0^2 = \frac{\sigma_{\rm v}^{2}}{R_{\rm eff}} = \frac{1}{3}\left(\frac{3\pi G \Sigma}{5} + \frac{4P_{\rm e}}{\Sigma}\right) .
\end{equation}
Within this framework, the locations of Galactic edge clumps in the
$\sigma_{\rm v}$--$R$ plane imply typical external pressures of
$P_{\rm e}$\,$\sim$\,$10^4$--$10^6$\,K\,cm$^{-3}$ (see Fig.\,\ref{fig:size-velocity dispersion}), 
comparable to the pressures required to confine outer-disk molecular regions \citep{Heyer2001}. 
Furthermore, external pressures in the Perseus arm ($\sim$\,10$^2$--10$^5$\,K\,cm$^{-3}$), 
Outer arm ($\sim$\,10$^2$--10$^4$\,K\,cm$^{-3}$), and OSC arm clouds ($\sim$\,10$^2$--10$^4$\,K\,cm$^{-3}$) 
are relatively minor compared to our Galactic edge clumps (see Fig.\,\ref{fig:size-velocity dispersion}). 
By contrast, much higher pressures are inferred for extreme, feedback-dominated environments such 
as the Galactic centre and 30\,Dor \citep{Oka1998,Oka2001,Indebetouw2013,Lu2024,Veena2024,Zhang2025}, 
as well as for dense gas in interacting systems \citep{Krahm2024}. 
Thus, while the outer-Galaxy environment is relatively low-pressure, the inferred $P_{\rm e}$ 
appears sufficient to confine most Galactic edge clumps, highlighting the importance of pressure 
confinement in regulating molecular structures at the Galactic edge. 

\begin{figure}[t]
\centering
\includegraphics[width=0.5\textwidth]{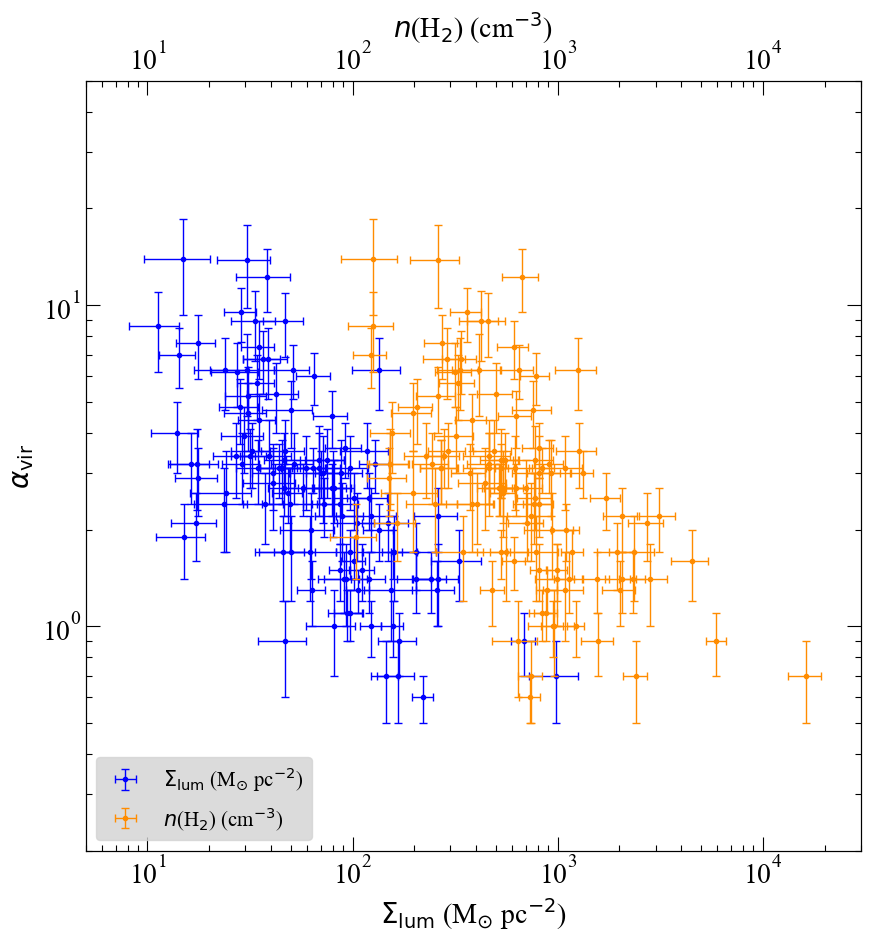}
\caption{Variation of the clump's virial parameter with surface density 
($\Sigma_{\rm lum}$, \emph{blue}) and volume density ($n$(H$_2$), \emph{orange}). 
} 
\label{fig:density-alpha} 
\end{figure}

The virial parameter displays a systematic decrease at $R_{\rm g}$\,$\sim$\,14--23\,kpc 
($\alpha_{\rm vir}$\,=\,33.0($\pm$\,10.4)\,e$^{-R_{\rm g}/6.7(\pm0.9)}$. 
The Pearson's coefficient is $-$0.61, indicating a strong negative correlation. 
And the $p$-value less than $10^{-3}$ indicates high significance (see Fig.\,\ref{fig:Rg-alpha}). 
Nevertheless, the clump's $\alpha_{\rm vir}$ exhibits larger values in our Galactic edge 
compared to those in the Outermost Galaxy, as observed through $^{13}$CO lines \citep{Lin2025}. 
In addition, the virial parameter is consistently larger in our Galactic edge clumps compared 
to those clouds in the OSC arm \citep{Sun2024a}. A plausible scenario is that an increase in 
the optical depths of CO\,(2--1) results in a larger velocity dispersion, consequently making 
the virial parameter also larger. Numerous studies have examined the relation between virial parameters
and masses of molecular clouds, denoted as $\alpha_{\rm vir}$\,=\,$iM_{\rm lum}^{j}$ 
(e.g., \citealt{Kauffmann2013,Urquhart2014,Miville-Deschenes2017,Ballesteros-Paredes2020,Li2023,Sun2024a}). 
The observed index $j$ was within the range of -1\,<\,$j$\,<\,0 \citep{Kauffmann2013}. 
The relation depicts a decline in the values of virial parameters as the clump mass increases. 
This indicates that the least gravitationally stable regions are located within the most massive clumps \citep{Urquhart2014}. 
The correlation between virial parameters and masses has been examined in our Galactic edge clumps ($\alpha_{\rm vir}$\,=\,27.24($\pm$7.43)$M_{\rm lum}^{\,-0.41\,(\pm0.05)}$. 
The Pearson's coefficient of $-$0.63 indicates a strong negative correlation, and the $p$-value less than $10^{-3}$ indicates high significance (see Fig.\,\ref{fig:Rg-alpha}). 
This correlation aligns with findings from the outer disk of the Milky Way ($\alpha_{\rm vir}$\,=\,37.2\,$M_{\rm lum}^{\,-0.40\,(\pm0.01)}$) 
as reported by \citet{Sun2024a}. 
Clouds with masses greater than $\sim$\,10$^4$\,M$_{\odot}$ may be undergoing gravitational collapse, whereas those with masses less than $\sim$\,10$^3$\,M$_{\odot}$ are unlikely to be gravitationally bound in the outer Galaxy. 
This finding is consistent with the results 
presented by \citet{Heyer2001}. Moreover, an anti-correlation is observed between the virial 
parameter and both volume and surface densities (see Fig.\,\ref{fig:density-alpha}), 
suggesting that clumps with higher volume and surface densities tend to be more gravitationally unstable. 
These findings suggest that instances of gravitational instability are more prevalent in clumps 
of greater mass and density. 

The threshold H$_2$ column density necessary for self-shielding is typically regarded as
$\sim$\,10$^{21}$\,cm$^{-2}$ (e.g., \citealt{Kirk2006,Jiao2025}), while star formation becomes 
prevalent when the column density surpasses 8\,$\times$\,10$^{21}$\,cm$^{-2}$ \citep{Vzquez-Semadeni2007}. 
In our Galactic edge clumps, the median column density is 2.9\,$\times$\,10$^{21}$\,cm$^{-2}$. 
This may suggest a lack of star formation within the clumps at the edge of the Galaxy. 
In addition, the volume densities of most clumps (see Table\,\ref{table:parameters} and 
Fig.\,\ref{fig:pra_hist}) are below the CO\,(2--1) critical density, which is 
on the order of 10$^3$\,cm$^{-3}$ for optical depths of about 10 \citep{Shirley2015,Penaloza2018}. 
Observed differences among molecular cloud and clump properties are jointly governed by: 
distance-related selection effects, angular resolution and beam dilution, tracer-dependent 
excitation and optical depth, large-scale Galactic environment (metallicity, pressure), 
and the star formation mode and evolutionary state (massive vs. low-mass; star-forming vs. quiescent). 

\subsection{CO-to-H$_2$ conversion factor}
\label{sect:conversion factor}
The $X_{\rm CO}$ factor exhibits variation across different environments 
(e.g., \citealt{Arimoto1996,Leroy2011,Bolatto2013,Amorin2016,Schruba2017,Saldano2023,Lee2024,Deng2025}). 
However, determination of the $X_{\rm CO}$ value is challenging. 
The assumption of a constant $X_{\rm CO}$ is frequently employed in the calculation of physical parameters for molecular clouds. 
The virial masses of the outer Galaxy clouds appear to be 2-3 times larger than those derived from the CO luminosity \citep{Sodroski1991}. 
It is more likely that $X_{\rm CO}$ is not constant but increases with Galactocentric radius due to a decrease in metallicity \citep{Abdo2010,Tibaldo2011}. 
Consequently, the luminous mass might be underestimated in the outer Galaxy, given that the $X_{\rm CO}$ factor is quantified in the Solar neighborhood. 
The virial parameter was examined to decrease with increasing Galactocentric radius (refer to Sect.\,\ref{sect:Stability of the clumps} or \citealt{Lin2025}). 
However, the virial parameter remains consistent in both inner and outer Galaxy, owing to the assumption of a constant $X_{\rm CO}$ factor \citep{Miville-Deschenes2017}. 
Furthermore, the decreased surface density of molecular clouds in the outer Galaxy may be partially attributed to a variation in the $X_{\rm CO}$ factor. 

\begin{figure}[t]
\centering
\includegraphics[width=0.5\textwidth]{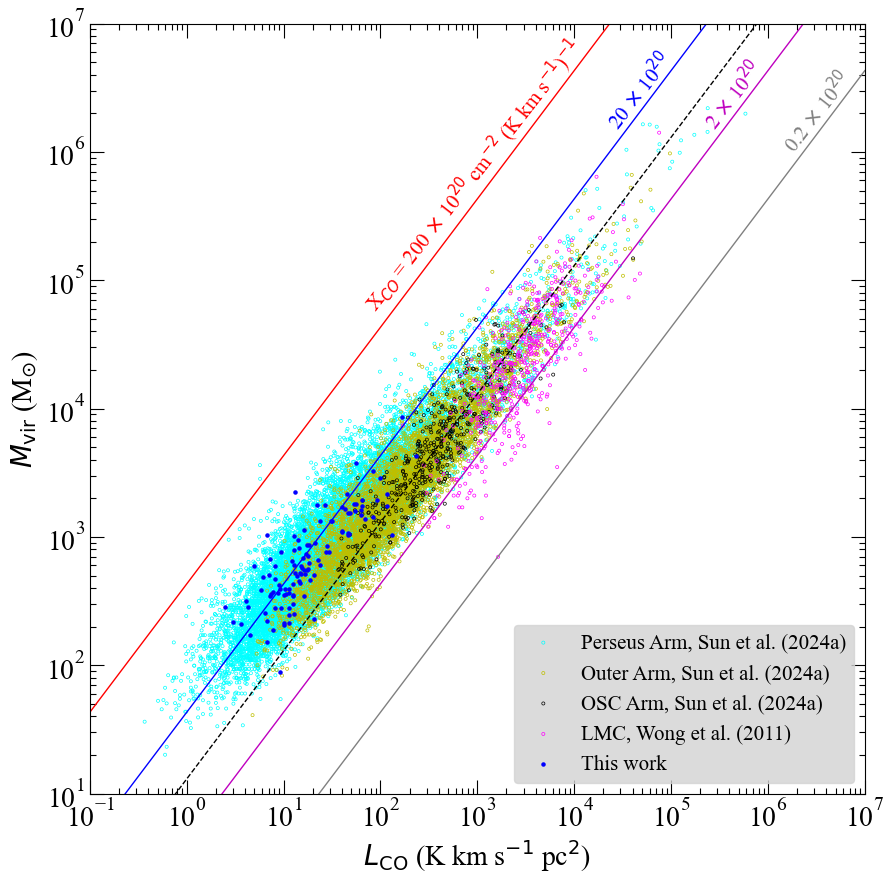}
\caption{Virial mass of molecular clouds as a function of CO luminosity. 
Results are obtained from molecular clouds located in the Perseus arm (\emph{cyan}), Outer arm (\emph{yellow}),
OSC arm (\emph{black}), Galactic edge (\emph{blue}), and LMC (\emph{magenta}). The dashed black line denotes the median value of
the $X_{\rm CO}$ factor ($\sim$\,6\,$\times$\,10$^{20}$\,cm$^{-2}$\,(K\,km s$^{-1}$)$^{-1}$, corresponding to
$\sim$14\,M$_{\odot}$\,(K\,km\,s$^{-1}$\,pc$^2$)$^{-1}$) used in this study (see Sect.\,\ref{sect:Estimation of Physical Parameters}). }
\label{fig:lum-Mvir} 
\end{figure}

In the absence of other reliable tracers, the $X_{\rm CO}$ factor can be estimated by a virial-based approach \citep{Arimoto1996}. 
The correlation between $M_{\rm vir}$ and $L_{\rm CO}$ for the Galactic edge clouds is depicted in Fig.\,\ref{fig:lum-Mvir}. 
The figure illustrates a one-to-one correspondence between the $X_{\rm CO}$ and $\alpha_{\rm CO}$ values.
A mean of $X_{\rm CO}$ factor within the Galactic edge clumps is 
20\,$\times$\,10$^{20}$\,cm$^{-2}$\,(K\,km\,s$^{-1}$)$^{-1}$ at $\sim$\,0.6\,pc resolution. 
However, the median $X_{\rm CO}$ factor is only 6\,$\times$\,10$^{20}$\,cm$^{-2}$\,(K\,km s$^{-1}$)$^{-1}$ 
for our Galactic edge clumps even with the relationship $X_{\rm CO}$\,$\propto$\,Z$^{-1.0}$ 
(see Sect.\,\ref{sect:Estimation of Physical Parameters}). 
Thus, the $X_{\rm CO}$ factor for these clumps appears to be underestimated. 
Nevertheless, the $X_{\rm CO}$ factor estimated by a virial-based approach depends on the virial parameter. 
A more plausible hypothesis could suggest that the clouds do not represent an unusually 
high $\alpha_{\rm CO}$ but rather are gravitationally unbound. 
The mean $\alpha_{\rm CO}$ value in our Galactic edge clumps is approximately 
43\,M$_{\odot}$\,(K\,km\,s$^{-1}$\,pc$^2$)$^{-1}$, which is ten times greater than those found 
in the Solar neighborhood (4.3\,M$_{\odot}$\,(K\,km\,s$^{-1}$\,pc$^2$)$^{-1}$). 
Similarly, high $\alpha_{\rm CO}$ values (10--28\,M$_{\odot}$\,(K\,km\,s$^{-1}$\,pc$^2$)$^{-1}$) 
are observed in metal-poor galaxies (e.g., SMC; \citealt{Saldano2023}). 
The $\alpha_{\rm CO}$ values show a clear increasing trend with greater Galactocentric distance \citep{Urquhart2024} 
and lower metallicity \citep{Wilson1995,Arimoto1996,Amorin2016,Bolatto2013}. 
This is comparable to the $X_{\rm CO}$ factor as determined through gamma-ray observations \citep{Abdo2010,Tibaldo2011,Luo2024c}. 
The equation of $X_{\rm CO}$\,$\propto$\,$Z^{-n}$ (n=1--3) is frequently examined in previous studies 
\citep{Arimoto1996,Bolatto2013,Amorin2016,Hunt2020,Hunt2023}. 
The outer Galaxy resembles the early environment of galaxy formation, that is, the metallicity decreases 
with the distance from the Galactic center \citep{Martig2016}. 
In low-metal environments, CO emission is often markedly faint due to the diminished abundances of C and O. 
Furthermore, CO can be readily photodissociated by ultraviolet radiation due to a lack of obscuring dust \citep{ONeill2022a}. 

The $X$-factor in low-luminosity molecular clouds is larger than that in high-luminosity ones (see Fig.\,\ref{fig:lum-Mvir}). 
Previous observations suggest that $\alpha_{\rm CO}$ rises with a decrease in density \citep{Bolatto2013}. 
The CO-dark H$_2$ gas is characterized by a broad diffuse envelope of H$_2$ gas that either does not or only weakly emits CO. As mentioned in Sect.\,\ref{sect:Introduction}, 
prior studies suggest that the fraction of CO-dark H$_2$ to total H$_2$ ranges from $\sim$20\% 
at a Galactocentric distance of 4\,kpc to $\sim$80\% at 10\,kpc \citep{Pineda2013}. 
In the Solar neighborhood, CO-dark H$_2$ gas constitutes approximately half of the total molecular gas 
(e.g., \citealt{Paradis2012,Pineda2013,Chen2015}). 
This contribution may be even more pronounced within the regions of low metallicity present
at the Galactic edge \citep{Pineda2013,Langer2014}. 
As a result, only regions with particularly high column density are seen in CO emission, 
thereby enhancing the $\alpha_{\rm CO}$ value. CO line intensities are reduced when the emitting region is not fully resolved. 
Therefore, the $X_{\rm CO}$ factors also depend on the spatial resolution \citep{Rubio1991,Rubio1993,Henkel2022}. 
The mean $X_{\rm CO}$ factors for the Perseus arm, Outer arm, OSC arm clouds 
($\theta_{\rm beam}$\,$\sim$\,50$^{\prime\prime}$; \citealt{Sun2024a}), and the LMC clouds 
($\theta_{\rm beam}$\,$\sim$\,45$^{\prime\prime}$; \citealt{Wong2011}) are 16, 8, 7, 
and 4\,$\times$\,10$^{20}$\,cm$^{-2}$\,(K\,km\,s$^{-1}$)$^{-1}$, respectively (see Fig.\,\ref{fig:lum-Mvir}). 
It is important to note that the physical parameters of our Galactic edge clumps are determined by 
CO\,(2--1), while other samples are derived from CO\,(1--0). 
The $X_{\rm CO}$ factors in our Galactic edge clumps closely align with those clouds found in the Perseus arm. 
In addition, the typical spatial resolution of our Galactic edge clumps ($\sim$\,0.6\,pc) exceeds 
that of the Perseus arm ($\sim$\,1\,pc), the Outer arm clouds ($\sim$\,3\,pc), the OSC clouds ($\sim$\,5\,pc), 
and the LMC clouds ($\sim$\,11\,pc). It is possible that this conversion factor varies on different scales, 
such as molecular cloud, clump, and core scales. 

\section{Summary}
\label{sect:summary}
We carried out CO\,(2--1) mapping observations, with an angular resolution of $\sim$11$^{\prime\prime}$,
towards 72 Galactic edge clouds utilizing the IRAM\,30\,m telescope. Based on these data, 
we investigated cloud-scale morphologies and derived the physical properties and scaling relations 
of molecular clumps within these clouds. 
The main results are the following: 
\begin{enumerate}
\item
Among the observed clouds, 25 exhibit compact formations, 25 manifest diffuse configurations, 
and 22 demonstrate an intermediate morphology bridging the characteristics of the aforementioned categories.

\item
A total of 112 CO clumps have been discerned within our observed Galactic edge clouds. 
The parameters of size, mass, surface density, and velocity dispersion of these clumps, 
as derived from CO observations, exhibit ranges of 0.6--3.4\,pc, 34--8250\,M$_\odot$, 
12--1025\,M$_{\odot}$\,pc$^{-2}$, and 0.3--1.7\,km\,s$^{-1}$, respectively.

\item
The size, velocity dispersion, surface density, and CO luminosity of Galactic edge clouds, 
as determined from CO\,(2--1) emission, show no discernible systematic variation across 
the Galactocentric distance range of 14 to 23\,kpc. However, the virial parameters display a decline described by an exponential relation $\alpha_{\rm vir}$\,=\,33.0($\pm$\,10.4)\,e$^{-R_{\rm g}/6.7(\pm0.9)}$ from $R_{\rm g}$\,=\,14 to 23\,kpc in the outer Galaxy, conforming previous observational results. 

\item
The velocity dispersion-size relation of the Galactic edge clouds is fitted as
$\sigma_{\rm v}$\,=\,0.69($\pm$0.03)$R_{\rm eff}^{0.36(\pm0.10)}$, suggesting that turbulent motions are present in the Galactic edge clumps. 
This finding mirrors that observed in inner Galactic disk clouds. 
It hints at a potential for star-formation activity within molecular clouds located at the Galactic edge.

\item
The mass-size relation of the Galactic edge clouds is fitted as 
$M_{\rm lum}$\,=\,196($\pm$17)$R_{\rm eff}^{\,2.18\,(\pm0.26)}$. 
This indicates that the clumps maintain a roughly constant average column density ($\sim$\,2.9\,$\times$\,10$^{21}$\,cm$^{-2}$) that is nearly independent of clump size for our Galactic edge cloud sample. 

\item
The virial parameters within the Galactic edge clumps range from 0.6 to 15.3,
with a mean of 3.4\,$\pm$\,0.8 and a median of 2.8\,$\pm$\,0.6. 
The virial parameters for the majority of clumps surpass 2, indicating that the 
clumps within the Galactic edge clouds are likely unbound. 
Nevertheless, the external pressure required to bind the clumps appears to be low 
at the edge of the Galaxy. 
\end{enumerate}

The outer Galaxy presents metal-poor environments conducive to exploring the physical 
properties and scaling relations of molecular clouds. It is crucial to conduct further 
observations of additional molecular lines, expand the cloud sample size, 
and improve resolution to facilitate a comprehensive investigation.

\begin{acknowledgements}
The authors thank the anonymous referee for helpful comments. 
We thank the staff of the IRAM telescope for their assistance in observations. 
This work acknowledges the support of the National Key R\&D Program of China 
under grant Nos.\,2023YFA1608002 and 2022YFA1603103, the Chinese Academy of Sciences 
(CAS) “Light of West China” Program under grant No.\,xbzg-zdsys-202212, 
and the Tianshan Talent Training Program of Xinjiang Uygur Autonomous Region under grant Nos.\,2022TSYCLJ0005. 
It was also partially supported by the Tianshan Talent Training Program of 
Xinjiang Uygur Autonomous Region under grant No.\,2024TSYCTD0013, 
the National Natural Science Foundation of China under grant Nos.\,12173075, 12373029, 12403033, and 12463006, 
the Xinjiang Key Laboratory of Radio Astrophysics under grant No.\,2023D04033, 
the Natural Science Foundation of Xinjiang Uygur Autonomous Region 
under grant Nos.\,2022D01A359 and 2023D01A11, the Central Guidance for Local Science
and Technology Development Fund under grant No.\,ZYYD2025ZY23, and the Youth Innovation 
Promotion Association CAS.  C.\,Henkel acknowledges support by the Chinese Academy of 
Sciences President's International Fellowship Initiative under grant No.\,2025PVA0048. 
T.\,Liu, X.\,P.\,Chen, K.\,Wang, J.\,W.\,Wu, and G.\,Wu acknowledge support by the 
Tianchi Talent Program of Xinjiang Uygur Autonomous Region. 
This research has used NASA's Astrophysical Data System (ADS). 
\end{acknowledgements}

\bibliography{larson} 

\begin{appendix}
\onecolumn

\section{CO velocity-integrated intensity maps of the Galactic edge clouds}
\label{Sect:Int-maps2}
\setcounter{figure}{0}
\begin{figure*}[h]
\centering
\includegraphics[width=0.96\textwidth]{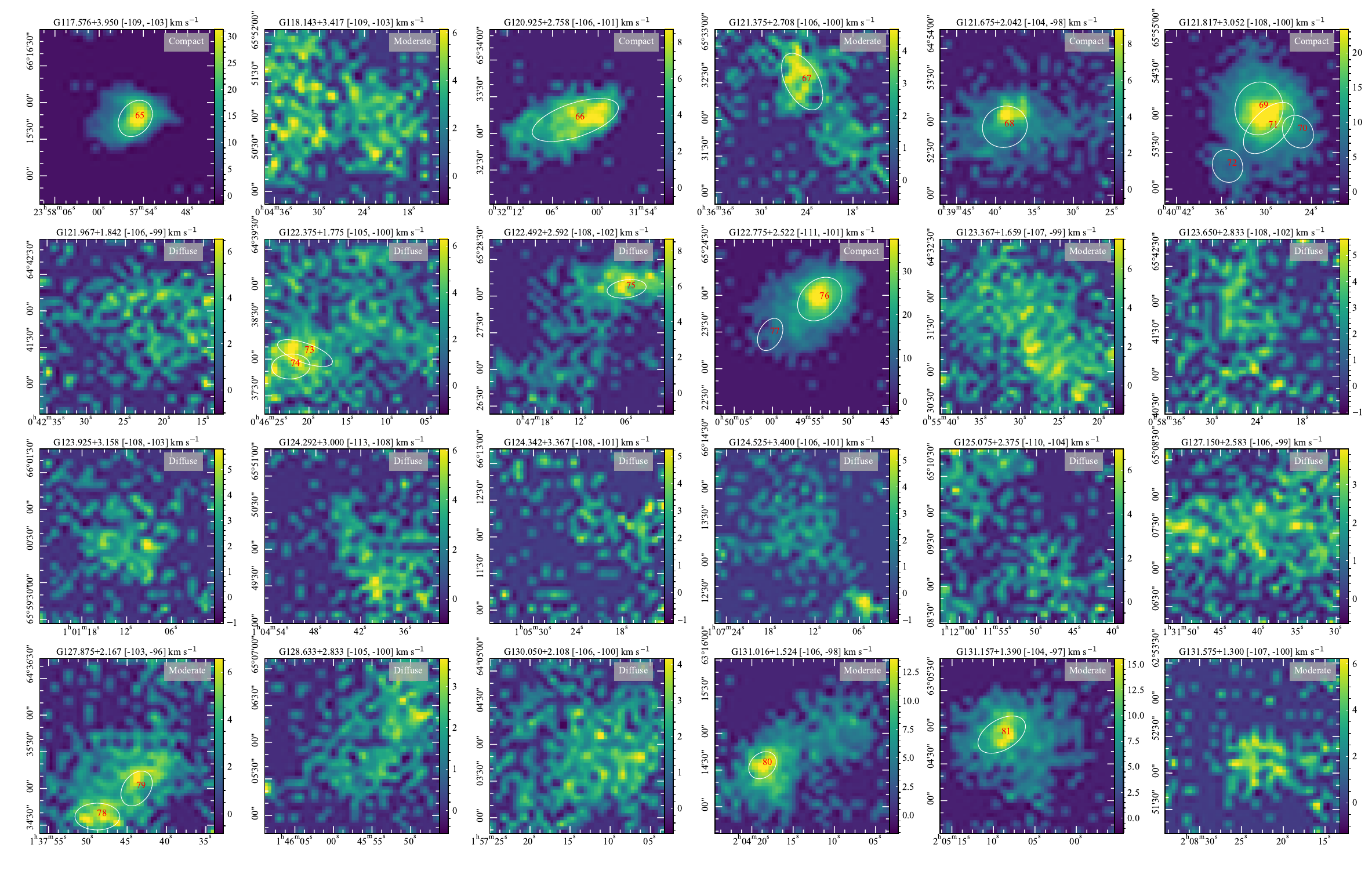}

\vspace{-4.05mm}

\includegraphics[width=0.96\textwidth]{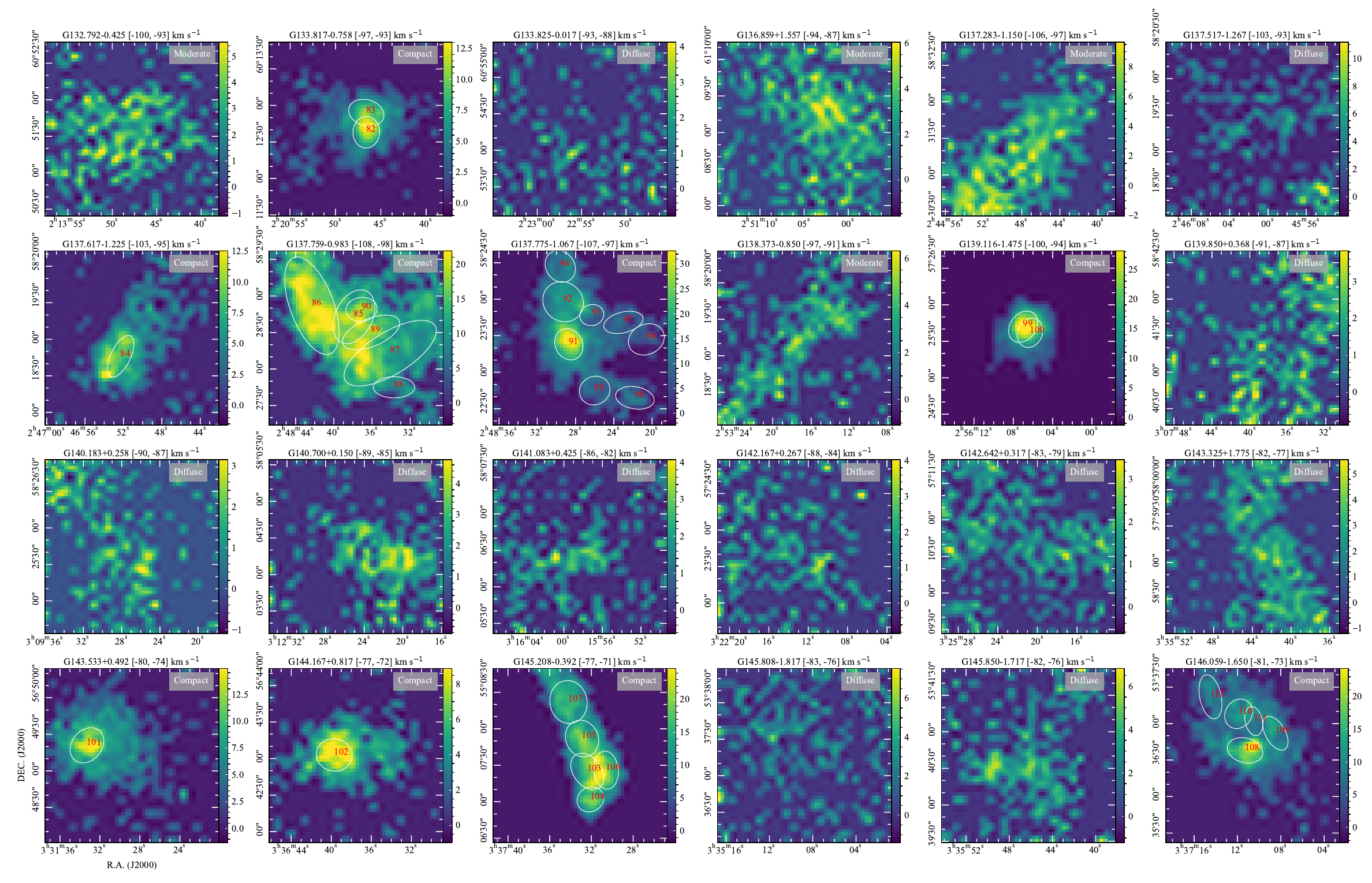}
\caption{CO\,(2--1) velocity-integrated intensity maps of the Galactic edge clouds. 
The detailed descriptions are the same as in Fig.\,\ref{fig:map}.}
\label{fig:map2} 
\end{figure*}

\section{CO channel maps of G145.208-0.392}
\label{Sect:channel maps}
\setcounter{figure}{0}
\begin{figure*}[h]
\centering
\includegraphics[width=1\textwidth]{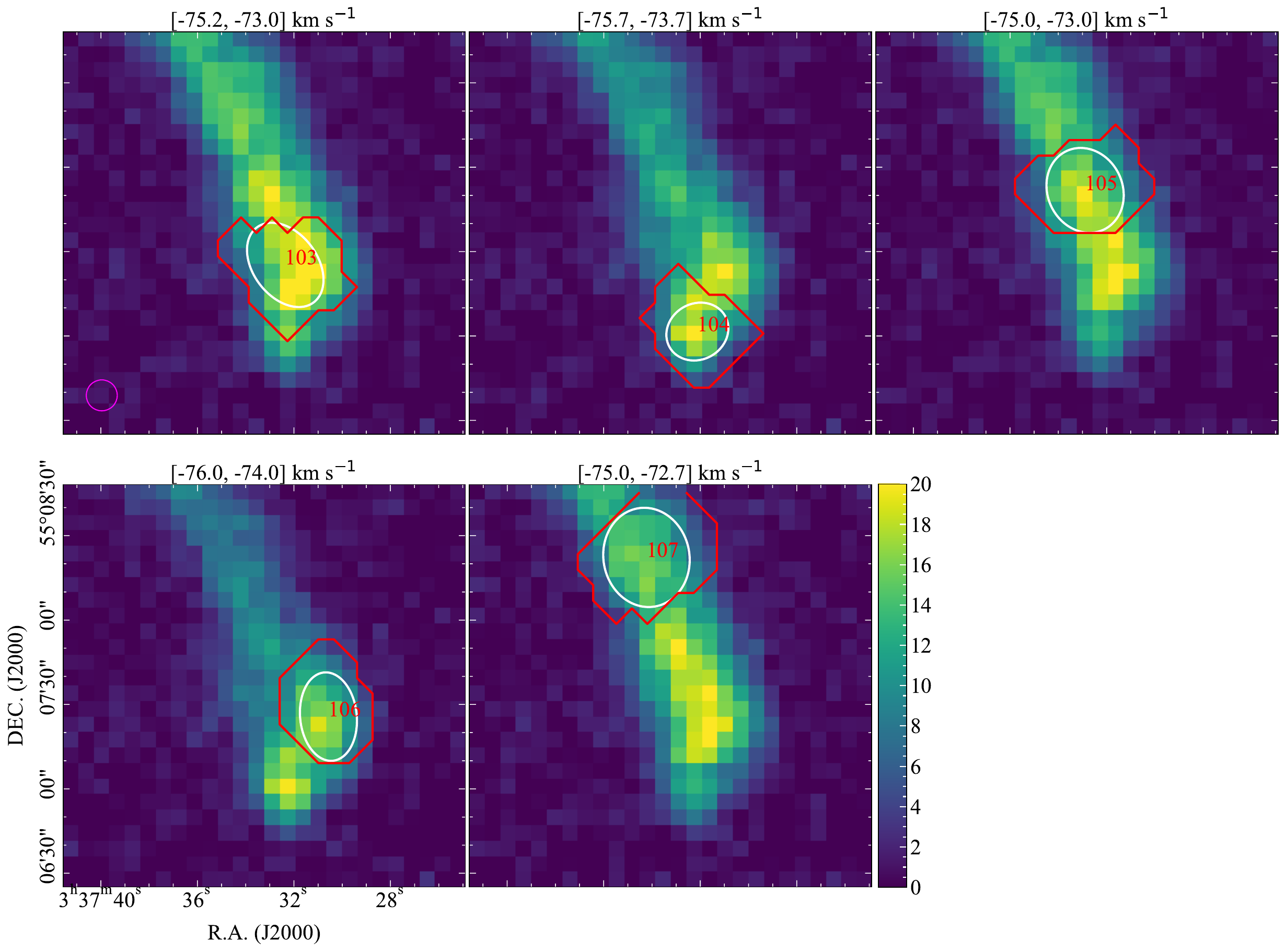}
\caption{Clumps recognized in 3D-PPV space at different velocities for G145.208-0.392 molecular clouds. 
The color maps represent the CO\,(2--1) velocity-integrated intensity on a $\int T_{\rm MB}dv$ scale, with the corresponding color bar (in units of K\,km\,s$^{-1}$) shown to the right of the final panel. 
The velocity-integration range indicated at the top of each panel corresponds to the slices in which clumps are identified. 
Red segments and white ellipses represent the identified and fitted CO clumps (see Sect.\,\ref{sect:Overview}), respectively, numbered from 103 to 107 as presented in Table\,\ref{table:parameters}.  
The magenta circle in the lower-left corner of the first panel shows the beam size of 
the CO\,(2--1) data obtained by the IRAM\,30\,m.} 
\label{fig:channel maps} 
\end{figure*}

\newpage
\onecolumn
\section{Physical parameters of CO clumps}
\label{Sect:Physical parameters of CO clumps}

\begingroup
\setlength{\tabcolsep}{3.6pt}
\setcounter{table}{0}
\begin{table*}[h]
\scriptsize
\caption{Physical parameters of CO clumps.}
\centering
\begin{tabular}
{ccccccccccccccc}
\hline\hline
Num &R.A.(J2000) &Dec.(J2000) &$R_{\rm g}$ &$R_{\rm eff}$ &$V_{\rm LSR}$ &$\sigma_{\rm v}$ &$T_{\rm mb}$ &$I_{\rm CO}$ &$L_{\rm CO}$ &$M_{\rm lum}$ &$N(\rm H_2)$ &$n(\rm H_2)$ &$\alpha_{\rm vir}$\\
&(h m s) &($\circ$ $\prime$ $\prime\prime$) &(kpc) &(pc) &(km\,s$^{-1}$) &(km\,s$^{-1}$) &(K) &(K\,km\,s$^{-1}$) &(K\,km\,s$^{-1}$\,pc$^{2}$) &(M$_{\odot}$) &($10^{21}$\,cm$^{-2}$) &($10^{2}$\,cm$^{-3}$) & \\
(1) &(2) &(3) &(4) &(5) &(6) &(7) &(8) &(9) &(10) &(11) &(12) &(13) &(14)\\
\hline
1   &21:58:11.2 &58:38:56  &14.4\,$\pm$\,0.2  &1.66\,$\pm$\,0.04  &-100.1  &0.99\,$\pm$\,0.04  &3.6\,$\pm$\,0.1  &9.1 \,$\pm$\,0.3  &62.1 \,$\pm$\,8.1   &597 \,$\pm$\,123   &3.2 \,$\pm$\,0.7   &4.6 \,$\pm$\,1.0  &3.2 \,$\pm$\,0.8\\
2   &21:58:05.8 &58:38:53  &14.4\,$\pm$\,0.2  &1.36\,$\pm$\,0.03  &-100.7  &0.76\,$\pm$\,0.03  &3.4\,$\pm$\,0.1  &6.5 \,$\pm$\,0.3  &19.3 \,$\pm$\,2.5   &185 \,$\pm$\,38    &1.5 \,$\pm$\,0.3   &2.6 \,$\pm$\,0.5  &4.9 \,$\pm$\,1.1\\
3   &21:58:07.2 &58:39:50  &14.4\,$\pm$\,0.2  &1.57\,$\pm$\,0.04  &-101.7  &1.02\,$\pm$\,0.04  &3.5\,$\pm$\,0.1  &9.0 \,$\pm$\,0.4  &59.3 \,$\pm$\,7.7   &570 \,$\pm$\,117   &3.4 \,$\pm$\,0.7   &5.2 \,$\pm$\,1.1  &3.3 \,$\pm$\,0.8\\
4   &21:58:02.9 &58:38:35  &14.4\,$\pm$\,0.2  &1.02\,$\pm$\,0.03  &-100.6  &0.50\,$\pm$\,0.02  &3.6\,$\pm$\,0.1  &4.6 \,$\pm$\,0.2  &12.4 \,$\pm$\,1.7   &120 \,$\pm$\,25    &1.7 \,$\pm$\,0.4   &4.0 \,$\pm$\,0.8  &2.5 \,$\pm$\,0.6\\
5   &21:58:09.2 &58:38:23  &14.4\,$\pm$\,0.2  &1.02\,$\pm$\,0.03  &-102.7  &0.57\,$\pm$\,0.02  &3.3\,$\pm$\,0.1  &4.8 \,$\pm$\,0.2  &9.7  \,$\pm$\,1.3   &93.3\,$\pm$\,19.2  &1.3 \,$\pm$\,0.3   &3.1 \,$\pm$\,0.7  &4.1 \,$\pm$\,1.0\\
6   &21:58:06.4 &58:38:57  &14.4\,$\pm$\,0.2  &1.26\,$\pm$\,0.03  &-102.6  &0.86\,$\pm$\,0.04  &2.8\,$\pm$\,0.1  &6.1 \,$\pm$\,0.3  &7.4  \,$\pm$\,1.0   &70.8\,$\pm$\,14.6  &0.7 \,$\pm$\,0.1   &1.3 \,$\pm$\,0.3  &15.3\,$\pm$\,3.5\\
7   &21:59:50.8 &58:31:16  &14.4\,$\pm$\,0.2  &1.19\,$\pm$\,0.02  &-102.3  &1.14\,$\pm$\,0.04  &4.1\,$\pm$\,0.1  &11.7\,$\pm$\,0.5  &22.7 \,$\pm$\,2.4   &219 \,$\pm$\,42    &2.3 \,$\pm$\,0.4   &4.6 \,$\pm$\,0.9  &8.2 \,$\pm$\,1.7\\
8   &21:59:52.2 &58:31:17  &14.4\,$\pm$\,0.2  &1.16\,$\pm$\,0.02  &-103.3  &0.79\,$\pm$\,0.04  &3.9\,$\pm$\,0.1  &7.7 \,$\pm$\,0.4  &14.9 \,$\pm$\,1.6   &144 \,$\pm$\,27    &1.6 \,$\pm$\,0.3   &3.3 \,$\pm$\,0.6  &5.8 \,$\pm$\,1.3\\
9   &21:59:53.4 &58:31:47  &14.4\,$\pm$\,0.2  &1.05\,$\pm$\,0.02  &-103.1  &0.69\,$\pm$\,0.03  &3.4\,$\pm$\,0.1  &5.9 \,$\pm$\,0.3  &16.2 \,$\pm$\,1.7   &156 \,$\pm$\,30    &2.1 \,$\pm$\,0.4   &4.8 \,$\pm$\,0.9  &3.7 \,$\pm$\,0.8\\
10  &21:59:46.8 &58:31:11  &14.4\,$\pm$\,0.2  &0.92\,$\pm$\,0.02  &-102.8  &0.85\,$\pm$\,0.04  &3.5\,$\pm$\,0.1  &7.6 \,$\pm$\,0.3  &9.7  \,$\pm$\,1.1   &93.6\,$\pm$\,17.9  &1.6 \,$\pm$\,0.3   &4.3 \,$\pm$\,0.8  &8.2 \,$\pm$\,1.8\\
11  &21:59:57.3 &58:31:18  &14.4\,$\pm$\,0.2  &0.77\,$\pm$\,0.02  &-103.3  &0.57\,$\pm$\,0.03  &3.3\,$\pm$\,0.1  &4.8 \,$\pm$\,0.2  &4.4  \,$\pm$\,0.5   &42.8\,$\pm$\,8.2   &1.1 \,$\pm$\,0.2   &3.3 \,$\pm$\,0.6  &6.8 \,$\pm$\,1.5\\
12  &22:01:50.2 &58:40:01  &14.5\,$\pm$\,0.3  &1.30\,$\pm$\,0.04  &-100.5  &0.84\,$\pm$\,0.02  &5.7\,$\pm$\,0.1  &12.2\,$\pm$\,0.3  &49.4 \,$\pm$\,7.2   &480 \,$\pm$\,104   &4.2 \,$\pm$\,1.0   &7.8 \,$\pm$\,1.7  &2.2 \,$\pm$\,0.5\\
13  &22:01:46.6 &58:39:12  &14.5\,$\pm$\,0.3  &1.27\,$\pm$\,0.04  &-100.9  &0.62\,$\pm$\,0.02  &4.5\,$\pm$\,0.1  &7.1 \,$\pm$\,0.3  &17.4 \,$\pm$\,2.6   &169 \,$\pm$\,37    &1.6 \,$\pm$\,0.4   &2.9 \,$\pm$\,0.6  &3.3 \,$\pm$\,0.8\\
14  &22:01:43.6 &58:40:06  &14.5\,$\pm$\,0.3  &1.02\,$\pm$\,0.04  &-100.5  &0.39\,$\pm$\,0.02  &3.9\,$\pm$\,0.1  &4.0 \,$\pm$\,0.2  &4.8  \,$\pm$\,0.7   &46.5\,$\pm$\,10.1  &0.7 \,$\pm$\,0.2   &1.6 \,$\pm$\,0.3  &3.9 \,$\pm$\,1.0\\
15  &22:07:08.9 &58:50:33  &14.9\,$\pm$\,0.3  &1.42\,$\pm$\,0.05  &-106.9  &0.69\,$\pm$\,0.03  &2.6\,$\pm$\,0.1  &4.6 \,$\pm$\,0.2  &30.5 \,$\pm$\,6.7   &311 \,$\pm$\,83    &2.3 \,$\pm$\,0.6   &3.9 \,$\pm$\,1.0  &2.5 \,$\pm$\,0.8\\
16  &22:06:00.6 &59:46:01  &15.2\,$\pm$\,0.5  &0.97\,$\pm$\,0.06  &-108.3  &0.46\,$\pm$\,0.02  &4.1\,$\pm$\,0.1  &4.9 \,$\pm$\,0.2  &22.4 \,$\pm$\,6.0   &240 \,$\pm$\,73    &3.8 \,$\pm$\,1.4   &9.3 \,$\pm$\,2.9  &1.0 \,$\pm$\,0.4\\
17  &22:05:57.8 &59:45:56  &15.2\,$\pm$\,0.5  &0.83\,$\pm$\,0.05  &-108.3  &0.31\,$\pm$\,0.01  &4.0\,$\pm$\,0.1  &3.2 \,$\pm$\,0.1  &9.8  \,$\pm$\,2.7   &105 \,$\pm$\,32    &2.3 \,$\pm$\,0.9   &6.6 \,$\pm$\,2.0  &0.9 \,$\pm$\,0.3\\
18  &22:05:55.1 &59:46:18  &15.2\,$\pm$\,0.5  &0.98\,$\pm$\,0.06  &-108.2  &0.48\,$\pm$\,0.02  &3.6\,$\pm$\,0.1  &4.4 \,$\pm$\,0.2  &14.0 \,$\pm$\,3.8   &150 \,$\pm$\,46    &2.3 \,$\pm$\,0.8   &5.7 \,$\pm$\,1.7  &1.7 \,$\pm$\,0.6\\
19  &22:09:42.6 &59:34:10  &14.5\,$\pm$\,0.5  &1.52\,$\pm$\,0.09  &-100.5  &1.03\,$\pm$\,0.03  &4.6\,$\pm$\,0.1  &12.0\,$\pm$\,0.4  &29.2 \,$\pm$\,6.5   &283 \,$\pm$\,78    &1.8 \,$\pm$\,0.5   &2.9 \,$\pm$\,0.8  &6.6 \,$\pm$\,2.0\\
20  &22:09:43.6 &59:33:15  &14.5\,$\pm$\,0.5  &0.81\,$\pm$\,0.05  &-101.4  &0.66\,$\pm$\,0.03  &4.1\,$\pm$\,0.1  &6.8 \,$\pm$\,0.3  &6.4  \,$\pm$\,1.5   &62.0\,$\pm$\,17.1  &1.4 \,$\pm$\,0.5   &4.2 \,$\pm$\,1.2  &6.6 \,$\pm$\,2.0\\
21  &22:15:56.6 &60:38:45  &14.8\,$\pm$\,0.4  &0.94\,$\pm$\,0.04  &-103.2  &1.20\,$\pm$\,0.03  &6.0\,$\pm$\,0.1  &18.1\,$\pm$\,0.5  &71.2 \,$\pm$\,8.9   &725 \,$\pm$\,144   &12.2\,$\pm$\,2.9   &31.0\,$\pm$\,6.1  &2.2 \,$\pm$\,0.5\\
22  &22:16:00.0 &60:38:45  &14.8\,$\pm$\,0.4  &1.49\,$\pm$\,0.06  &-102.7  &1.00\,$\pm$\,0.03  &6.0\,$\pm$\,0.1  &15.0\,$\pm$\,0.4  &71.6 \,$\pm$\,8.9   &729 \,$\pm$\,144   &4.9 \,$\pm$\,1.0   &7.8 \,$\pm$\,1.6  &2.4 \,$\pm$\,0.5\\
23  &22:15:59.5 &60:38:00  &14.8\,$\pm$\,0.4  &1.29\,$\pm$\,0.05  &-102.6  &0.61\,$\pm$\,0.02  &4.3\,$\pm$\,0.1  &6.7 \,$\pm$\,0.2  &16.7 \,$\pm$\,2.1   &170 \,$\pm$\,34    &1.5 \,$\pm$\,0.3   &2.8 \,$\pm$\,0.6  &3.3 \,$\pm$\,0.7\\
24  &22:19:07.9 &60:32:57  &14.2\,$\pm$\,0.3  &1.41\,$\pm$\,0.05  &-100.7  &0.64\,$\pm$\,0.02  &6.1\,$\pm$\,0.1  &9.9 \,$\pm$\,0.2  &22.9 \,$\pm$\,2.9   &213 \,$\pm$\,44    &1.6 \,$\pm$\,0.3   &2.7 \,$\pm$\,0.6  &3.1 \,$\pm$\,0.7\\
25  &22:18:57.9 &60:32:50  &14.2\,$\pm$\,0.3  &1.84\,$\pm$\,0.06  &-101.0  &0.85\,$\pm$\,0.03  &5.5\,$\pm$\,0.1  &11.8\,$\pm$\,0.3  &49.6 \,$\pm$\,6.2   &463 \,$\pm$\,96    &2.0 \,$\pm$\,0.4   &2.6 \,$\pm$\,0.6  &3.3 \,$\pm$\,0.8\\
26  &22:18:33.2 &60:41:55  &15.0\,$\pm$\,0.2  &1.19\,$\pm$\,0.03  &-106.9  &1.66\,$\pm$\,0.05  &6.1\,$\pm$\,0.1  &25.4\,$\pm$\,0.8  &58.3 \,$\pm$\,6.6   &610 \,$\pm$\,113   &6.4 \,$\pm$\,1.3   &12.9\,$\pm$\,2.4  &6.2 \,$\pm$\,1.3\\
27  &22:50:33.4 &61:52:12  &15.0\,$\pm$\,0.4  &1.17\,$\pm$\,0.06  &-103.0  &0.34\,$\pm$\,0.01  &4.4\,$\pm$\,0.1  &4.0 \,$\pm$\,0.1  &7.3  \,$\pm$\,1.4   &75.5\,$\pm$\,18.1  &0.8 \,$\pm$\,0.2   &1.7 \,$\pm$\,0.4  &2.1 \,$\pm$\,0.6\\
28  &22:50:35.3 &61:52:41  &15.0\,$\pm$\,0.4  &1.56\,$\pm$\,0.08  &-103.0  &0.35\,$\pm$\,0.01  &4.5\,$\pm$\,0.1  &4.1 \,$\pm$\,0.1  &10.8 \,$\pm$\,2.0   &112 \,$\pm$\,27    &0.7 \,$\pm$\,0.2   &1.0 \,$\pm$\,0.3  &2.0 \,$\pm$\,0.5\\
29  &22:50:33.6 &61:53:12  &15.0\,$\pm$\,0.4  &1.35\,$\pm$\,0.07  &-103.0  &0.43\,$\pm$\,0.02  &3.9\,$\pm$\,0.1  &4.3 \,$\pm$\,0.2  &10.0 \,$\pm$\,1.9   &104 \,$\pm$\,25    &0.8 \,$\pm$\,0.2   &1.5 \,$\pm$\,0.4  &2.8 \,$\pm$\,0.8\\
30  &22:50:30.3 &61:53:25  &15.0\,$\pm$\,0.4  &1.29\,$\pm$\,0.06  &-103.4  &0.44\,$\pm$\,0.02  &3.5\,$\pm$\,0.1  &3.9 \,$\pm$\,0.1  &8.7  \,$\pm$\,1.7   &90.7\,$\pm$\,21.7  &0.8 \,$\pm$\,0.2   &1.5 \,$\pm$\,0.4  &3.2 \,$\pm$\,0.9\\
31  &22:51:57.1 &61:44:53  &14.9\,$\pm$\,0.3  &1.43\,$\pm$\,0.05  &-101.1  &1.17\,$\pm$\,0.03  &4.3\,$\pm$\,0.1  &12.5\,$\pm$\,0.3  &50.8 \,$\pm$\,5.6   &520 \,$\pm$\,98    &3.8 \,$\pm$\,0.8   &6.3 \,$\pm$\,1.2  &4.4 \,$\pm$\,0.9\\
32  &22:51:58.0 &61:45:13  &14.9\,$\pm$\,0.3  &1.06\,$\pm$\,0.04  &-102.5  &0.73\,$\pm$\,0.04  &3.4\,$\pm$\,0.1  &6.2 \,$\pm$\,0.3  &20.4 \,$\pm$\,2.3   &209 \,$\pm$\,39    &2.8 \,$\pm$\,0.6   &6.2 \,$\pm$\,1.2  &3.1 \,$\pm$\,0.7\\
33  &22:51:58.1 &61:45:26  &14.9\,$\pm$\,0.3  &1.26\,$\pm$\,0.04  &-101.5  &0.91\,$\pm$\,0.04  &3.3\,$\pm$\,0.1  &7.5 \,$\pm$\,0.3  &18.3 \,$\pm$\,2.1   &187 \,$\pm$\,35    &1.7 \,$\pm$\,0.4   &3.3 \,$\pm$\,0.6  &6.5 \,$\pm$\,1.4\\
34  &22:51:58.8 &61:44:28  &14.9\,$\pm$\,0.3  &0.92\,$\pm$\,0.03  &-101.3  &0.88\,$\pm$\,0.03  &3.5\,$\pm$\,0.1  &7.7 \,$\pm$\,0.3  &14.1 \,$\pm$\,1.6   &145 \,$\pm$\,27    &2.5 \,$\pm$\,0.5   &6.6 \,$\pm$\,1.3  &5.7 \,$\pm$\,1.2\\
35  &22:51:55.9 &61:44:22  &14.9\,$\pm$\,0.3  &0.64\,$\pm$\,0.02  &-102.4  &0.67\,$\pm$\,0.03  &2.9\,$\pm$\,0.1  &4.9 \,$\pm$\,0.2  &4.5  \,$\pm$\,0.5   &46.3\,$\pm$\,8.7   &1.7 \,$\pm$\,0.4   &6.3 \,$\pm$\,1.2  &7.2 \,$\pm$\,1.6\\
36  &22:52:00.3 &61:44:00  &14.9\,$\pm$\,0.3  &0.94\,$\pm$\,0.03  &-100.4  &0.81\,$\pm$\,0.03  &3.2\,$\pm$\,0.1  &6.5 \,$\pm$\,0.3  &8.1  \,$\pm$\,0.9   &83.2\,$\pm$\,15.7  &1.4 \,$\pm$\,0.3   &3.6 \,$\pm$\,0.7  &8.6 \,$\pm$\,1.8\\
37  &22:51:57.1 &61:43:56  &14.9\,$\pm$\,0.3  &0.66\,$\pm$\,0.03  &-100.0  &0.90\,$\pm$\,0.03  &2.7\,$\pm$\,0.1  &6.1 \,$\pm$\,0.2  &5.4  \,$\pm$\,0.6   &55.6\,$\pm$\,10.5  &1.9 \,$\pm$\,0.5   &6.9 \,$\pm$\,1.3  &11.1\,$\pm$\,2.3\\
38  &22:52:03.6 &61:44:04  &14.9\,$\pm$\,0.3  &0.76\,$\pm$\,0.03  &-100.6  &0.51\,$\pm$\,0.03  &2.5\,$\pm$\,0.1  &3.2 \,$\pm$\,0.2  &3.3  \,$\pm$\,0.4   &34.2\,$\pm$\,6.5   &0.9 \,$\pm$\,0.2   &2.8 \,$\pm$\,0.5  &6.7 \,$\pm$\,1.6\\
39  &22:50:24.2 &62:16:56  &14.4\,$\pm$\,0.1  &1.17\,$\pm$\,0.02  &-98.5   &0.41\,$\pm$\,0.02  &3.3\,$\pm$\,0.1  &3.5 \,$\pm$\,0.2  &7.2  \,$\pm$\,0.9   &69.3\,$\pm$\,13.5  &0.8 \,$\pm$\,0.1   &1.5 \,$\pm$\,0.3  &3.3 \,$\pm$\,0.8\\
40  &22:50:32.9 &62:17:07  &14.4\,$\pm$\,0.1  &1.32\,$\pm$\,0.02  &-98.9   &0.84\,$\pm$\,0.04  &3.4\,$\pm$\,0.1  &7.3 \,$\pm$\,0.3  &49.9 \,$\pm$\,5.7   &481 \,$\pm$\,94    &4.1 \,$\pm$\,0.8   &7.4 \,$\pm$\,1.5  &2.2 \,$\pm$\,0.5\\
41  &22:51:39.0 &62:18:22  &14.7\,$\pm$\,0.3  &1.65\,$\pm$\,0.05  &-99.8   &0.98\,$\pm$\,0.04  &3.8\,$\pm$\,0.1  &9.3 \,$\pm$\,0.3  &71.6 \,$\pm$\,9.3   &712 \,$\pm$\,144   &3.9 \,$\pm$\,0.8   &5.6 \,$\pm$\,1.1  &2.6 \,$\pm$\,0.6\\
42  &22:51:31.1 &62:19:03  &14.7\,$\pm$\,0.3  &1.33\,$\pm$\,0.04  &-99.5   &0.76\,$\pm$\,0.03  &3.5\,$\pm$\,0.1  &6.7 \,$\pm$\,0.3  &30.7 \,$\pm$\,4.0   &305 \,$\pm$\,62    &2.6 \,$\pm$\,0.5   &4.6 \,$\pm$\,0.9  &2.9 \,$\pm$\,0.7\\
43  &22:51:29.8 &62:18:18  &14.7\,$\pm$\,0.3  &1.33\,$\pm$\,0.04  &-99.2   &0.58\,$\pm$\,0.03  &3.3\,$\pm$\,0.1  &4.9 \,$\pm$\,0.2  &15.0 \,$\pm$\,2.0   &149 \,$\pm$\,30    &1.3 \,$\pm$\,0.3   &2.3 \,$\pm$\,0.5  &3.5 \,$\pm$\,0.8\\
44  &22:51:36.8 &62:17:48  &14.7\,$\pm$\,0.3  &0.73\,$\pm$\,0.03  &-99.5   &0.68\,$\pm$\,0.03  &3.4\,$\pm$\,0.1  &5.8 \,$\pm$\,0.3  &8.6  \,$\pm$\,1.2   &85.4\,$\pm$\,17.3  &2.4 \,$\pm$\,0.6   &7.8 \,$\pm$\,1.6  &4.6 \,$\pm$\,1.1\\
45  &22:52:23.9 &62:27:22  &14.6\,$\pm$\,0.2  &1.71\,$\pm$\,0.04  &-98.9   &1.01\,$\pm$\,0.04  &4.5\,$\pm$\,0.1  &11.6\,$\pm$\,0.5  &102  \,$\pm$\,12    &1002\,$\pm$\,194   &5.1 \,$\pm$\,1.0   &7.1 \,$\pm$\,1.4  &2.0 \,$\pm$\,0.5\\
46  &22:52:15.8 &62:27:14  &14.6\,$\pm$\,0.2  &1.37\,$\pm$\,0.03  &-98.5   &0.58\,$\pm$\,0.03  &4.1\,$\pm$\,0.1  &6.1 \,$\pm$\,0.3  &18.3 \,$\pm$\,2.1   &180 \,$\pm$\,35    &1.4 \,$\pm$\,0.3   &2.5 \,$\pm$\,0.5  &3.0 \,$\pm$\,0.7\\
47  &22:53:27.8 &62:32:14  &14.6\,$\pm$\,0.3  &1.31\,$\pm$\,0.04  &-99.3   &0.87\,$\pm$\,0.03  &6.2\,$\pm$\,0.1  &13.8\,$\pm$\,0.5  &44.4 \,$\pm$\,5.0   &441 \,$\pm$\,84    &3.8 \,$\pm$\,0.8   &7.0 \,$\pm$\,1.3  &2.6 \,$\pm$\,0.6\\
48  &22:53:26.7 &62:32:02  &14.6\,$\pm$\,0.3  &0.61\,$\pm$\,0.02  &-100.1  &0.72\,$\pm$\,0.03  &5.6\,$\pm$\,0.1  &10.3\,$\pm$\,0.4  &17.9 \,$\pm$\,2.0   &178 \,$\pm$\,34    &7.1 \,$\pm$\,1.8   &27.8\,$\pm$\,5.3  &2.1 \,$\pm$\,0.5\\
49  &22:53:30.3 &62:32:50  &14.6\,$\pm$\,0.3  &1.24\,$\pm$\,0.04  &-99.3   &0.64\,$\pm$\,0.02  &5.5\,$\pm$\,0.1  &9.0 \,$\pm$\,0.3  &20.0 \,$\pm$\,2.3   &198 \,$\pm$\,38    &1.9 \,$\pm$\,0.4   &3.7 \,$\pm$\,0.7  &3.0 \,$\pm$\,0.7\\
50  &22:53:23.6 &62:32:35  &14.6\,$\pm$\,0.3  &1.32\,$\pm$\,0.04  &-99.2   &0.59\,$\pm$\,0.03  &4.4\,$\pm$\,0.1  &6.5 \,$\pm$\,0.3  &7.9  \,$\pm$\,0.9   &78.7\,$\pm$\,15.1  &0.7 \,$\pm$\,0.1   &1.2 \,$\pm$\,0.2  &6.8 \,$\pm$\,1.5\\
51  &22:53:46.4 &62:58:11  &14.3\,$\pm$\,0.2  &1.32\,$\pm$\,0.04  &-98.4   &0.48\,$\pm$\,0.02  &3.3\,$\pm$\,0.1  &4.1 \,$\pm$\,0.2  &13.2 \,$\pm$\,3.1   &125 \,$\pm$\,35    &1.1 \,$\pm$\,0.3   &1.9 \,$\pm$\,0.6  &2.8 \,$\pm$\,0.9\\
52  &22:53:53.6 &62:57:25  &14.3\,$\pm$\,0.2  &0.75\,$\pm$\,0.03  &-98.4   &0.50\,$\pm$\,0.02  &3.4\,$\pm$\,0.1  &4.4 \,$\pm$\,0.2  &11.8 \,$\pm$\,2.7   &112 \,$\pm$\,31    &3.0 \,$\pm$\,1.0   &9.4 \,$\pm$\,2.7  &1.9 \,$\pm$\,0.6\\
53  &22:53:49.9 &62:57:49  &14.3\,$\pm$\,0.2  &1.03\,$\pm$\,0.03  &-98.3   &0.40\,$\pm$\,0.02  &3.3\,$\pm$\,0.1  &3.5 \,$\pm$\,0.2  &8.2  \,$\pm$\,1.9   &77.5\,$\pm$\,21.7  &1.1 \,$\pm$\,0.3   &2.5 \,$\pm$\,0.7  &2.5 \,$\pm$\,0.8\\
54  &22:56:08.3 &62:45:07  &14.6\,$\pm$\,0.2  &1.03\,$\pm$\,0.02  &-98.7   &0.56\,$\pm$\,0.02  &3.9\,$\pm$\,0.1  &5.5 \,$\pm$\,0.2  &14.0 \,$\pm$\,2.6   &139 \,$\pm$\,34    &1.9 \,$\pm$\,0.5   &4.5 \,$\pm$\,1.1  &2.7 \,$\pm$\,0.7\\
55  &22:56:06.2 &62:45:26  &14.6\,$\pm$\,0.2  &0.97\,$\pm$\,0.02  &-98.3   &0.58\,$\pm$\,0.03  &3.3\,$\pm$\,0.1  &4.9 \,$\pm$\,0.2  &12.0 \,$\pm$\,2.3   &118 \,$\pm$\,29    &1.9 \,$\pm$\,0.5   &4.6 \,$\pm$\,1.1  &3.2 \,$\pm$\,0.9\\
56  &23:36:08.3 &62:23:53  &15.4\,$\pm$\,0.4  &0.71\,$\pm$\,0.03  &-101.2  &1.15\,$\pm$\,0.02  &9.8\,$\pm$\,0.1  &28.6\,$\pm$\,0.4  &148  \,$\pm$\,17    &1623\,$\pm$\,294   &47.7\,$\pm$\,12.1  &161 \,$\pm$\,30   &0.7 \,$\pm$\,0.2\\
57  &23:36:11.5 &62:23:42  &15.4\,$\pm$\,0.4  &1.37\,$\pm$\,0.06  &-100.8  &1.08\,$\pm$\,0.02  &8.0\,$\pm$\,0.1  &21.7\,$\pm$\,0.4  &131  \,$\pm$\,15    &1436\,$\pm$\,260   &11.3\,$\pm$\,2.3   &19.9\,$\pm$\,3.6  &1.3 \,$\pm$\,0.3\\
58  &23:36:15.4 &62:23:23  &15.4\,$\pm$\,0.4  &1.32\,$\pm$\,0.06  &-100.7  &1.07\,$\pm$\,0.03  &5.0\,$\pm$\,0.1  &13.4\,$\pm$\,0.4  &48.3 \,$\pm$\,5.4   &529 \,$\pm$\,96    &4.5 \,$\pm$\,0.9   &8.2 \,$\pm$\,1.5  &3.3 \,$\pm$\,0.7\\
59  &23:36:18.7 &62:24:20  &15.4\,$\pm$\,0.4  &1.76\,$\pm$\,0.08  &-101.4  &1.00\,$\pm$\,0.02  &5.5\,$\pm$\,0.1  &13.7\,$\pm$\,0.3  &73.2 \,$\pm$\,8.1   &803 \,$\pm$\,145   &3.8 \,$\pm$\,0.8   &5.2 \,$\pm$\,1.0  &2.5 \,$\pm$\,0.5\\
60  &23:50:38.6 &65:40:17  &16.4\,$\pm$\,0.4  &1.57\,$\pm$\,0.07  &-107.7  &0.91\,$\pm$\,0.03  &8.4\,$\pm$\,0.1  &19.2\,$\pm$\,0.5  &94.0 \,$\pm$\,12.2  &1176\,$\pm$\,213   &7.1 \,$\pm$\,1.4   &10.8\,$\pm$\,2.0  &1.3 \,$\pm$\,0.3\\
61  &23:50:39.8 &65:40:38  &16.4\,$\pm$\,0.4  &0.80\,$\pm$\,0.04  &-108.4  &0.79\,$\pm$\,0.03  &5.9\,$\pm$\,0.1  &11.6\,$\pm$\,0.5  &19.5 \,$\pm$\,2.6   &245 \,$\pm$\,44    &5.7 \,$\pm$\,1.5   &17.0\,$\pm$\,3.1  &2.4 \,$\pm$\,0.5\\
62  &23:50:32.7 &65:41:16  &16.4\,$\pm$\,0.4  &0.93\,$\pm$\,0.04  &-107.1  &0.62\,$\pm$\,0.02  &5.5\,$\pm$\,0.1  &8.7 \,$\pm$\,0.3  &10.3 \,$\pm$\,1.4   &128 \,$\pm$\,23    &2.2 \,$\pm$\,0.5   &5.7 \,$\pm$\,1.0  &3.2 \,$\pm$\,0.7\\
63  &23:50:45.4 &65:41:39  &16.4\,$\pm$\,0.4  &0.98\,$\pm$\,0.04  &-107.4  &0.67\,$\pm$\,0.03  &4.3\,$\pm$\,0.1  &7.4 \,$\pm$\,0.3  &6.8  \,$\pm$\,0.9   &84.7\,$\pm$\,15.4  &1.3 \,$\pm$\,0.3   &3.2 \,$\pm$\,0.6  &6.0 \,$\pm$\,1.3\\
64  &23:50:44.9 &65:41:14  &16.4\,$\pm$\,0.4  &1.01\,$\pm$\,0.04  &-107.9  &0.65\,$\pm$\,0.03  &4.6\,$\pm$\,0.1  &7.6 \,$\pm$\,0.3  &9.0  \,$\pm$\,1.2   &113 \,$\pm$\,20    &1.6 \,$\pm$\,0.4   &3.9 \,$\pm$\,0.7  &4.4 \,$\pm$\,1.0\\
65  &23:57:55.1 &66:15:47  &16.3\,$\pm$\,0.3  &1.06\,$\pm$\,0.04  &-106.0  &1.07\,$\pm$\,0.03  &7.3\,$\pm$\,0.1  &19.8\,$\pm$\,0.5  &34.5 \,$\pm$\,4.8   &423 \,$\pm$\,80    &5.6 \,$\pm$\,1.2   &12.6\,$\pm$\,2.4  &3.3 \,$\pm$\,0.7\\
\hline
\end{tabular}
\label{table:parameters}
\tablefoot{Column 1: numbers are consistent with the red numbers in Figs.\,\ref{fig:map} and \ref{fig:map2}. 
Columns 2--3: equatorial coordinates of the clump centroid derived from elliptical fitting. 
Columns 4--5: the Galactocentric distance ($R_{\rm g}$) and effective radius ($R_{\rm eff}$) of the clump. 
Columns 6--9: local standard of rest velocity ($V_{\rm LSR}$), velocity dispersion ($\sigma_{\rm v}$), 
main beam brightness temperature ($T_{\rm mb}$), and integrated intensity ($I_{\rm CO}$) obtained 
from Gaussian fitting from the averaged spectral profile over each clump. 
Columns 10--14: the CO luminosity ($L_{\rm CO}$), luminous masses ($M_{\rm lum}$), mean column 
density ($N(\rm H_2)$), mean volume density ($n(\rm H_2)$), and virial parameter ($\alpha_{\rm vir}$) 
of the clumps. For detailed derivation of parameters refer to Sect.\,\ref{sect:Estimation of Physical Parameters}. }
\end{table*}

\setcounter{table}{0}
\begin{table*}[h]
\scriptsize
\caption{continued.}
\centering
\begin{tabular}
{ccccccccccccccc}
\hline\hline
Num &R.A.(J2000) &Dec.(J2000) &$R_{\rm g}$ &$R_{\rm eff}$ &$V_{\rm LSR}$ &$\sigma_{\rm v}$ &$T_{\rm mb}$ &$I_{\rm CO}$ &$L_{\rm CO}$ &$M_{\rm lum}$ &$N(\rm H_2)$ &$n(\rm H_2)$ &$\alpha_{\rm vir}$\\
&(h m s) &($\circ$ $\prime$ $\prime\prime$) &(kpc) &(pc) &(km\,s$^{-1}$) &(km\,s$^{-1}$) &(K) &(K\,km\,s$^{-1}$) &(K\,km\,s$^{-1}$\,pc$^{2}$) &(M$_{\odot}$) &($10^{21}$\,cm$^{-2}$) &($10^{2}$\,cm$^{-3}$) & \\
(1) &(2) &(3) &(4) &(5) &(6) &(7) &(8) &(9) &(10) &(11) &(12) &(13) &(14)\\
\hline
66  &00:32:03.0 &65:33:11  &16.5\,$\pm$\,0.4  &1.82\,$\pm$\,0.07  &-102.7  &0.82\,$\pm$\,0.03  &3.7\,$\pm$\,0.1  &7.7 \,$\pm$\,0.3  &26.7 \,$\pm$\,4.8   &335 \,$\pm$\,73    &1.5 \,$\pm$\,0.3   &2.0 \,$\pm$\,0.4  &4.2 \,$\pm$\,1.0\\
67  &00:36:24.7 &65:32:31  &16.5\,$\pm$\,0.5  &1.45\,$\pm$\,0.08  &-103.6  &0.56\,$\pm$\,0.02  &3.0\,$\pm$\,0.1  &4.3 \,$\pm$\,0.2  &22.9 \,$\pm$\,6.5   &288 \,$\pm$\,89    &2.0 \,$\pm$\,0.7   &3.4 \,$\pm$\,1.0  &1.8 \,$\pm$\,0.6\\
68  &00:39:38.8 &64:52:56  &16.5\,$\pm$\,0.3  &1.37\,$\pm$\,0.04  &-101.3  &1.23\,$\pm$\,0.06  &2.6\,$\pm$\,0.1  &8.0 \,$\pm$\,0.4  &15.2 \,$\pm$\,3.1   &192 \,$\pm$\,46    &1.5 \,$\pm$\,0.4   &2.7 \,$\pm$\,0.6  &12.5\,$\pm$\,3.3\\
69  &00:40:31.0 &65:54:06  &16.5\,$\pm$\,0.2  &1.64\,$\pm$\,0.04  &-103.5  &1.33\,$\pm$\,0.04  &4.8\,$\pm$\,0.1  &16.1\,$\pm$\,0.4  &89.8 \,$\pm$\,7.7   &1137\,$\pm$\,169   &6.3 \,$\pm$\,1.0   &9.2 \,$\pm$\,1.4  &3.0 \,$\pm$\,0.5\\
70  &00:40:25.8 &65:53:47  &16.5\,$\pm$\,0.2  &0.93\,$\pm$\,0.03  &-104.0  &0.99\,$\pm$\,0.04  &3.6\,$\pm$\,0.1  &8.9 \,$\pm$\,0.3  &13.7 \,$\pm$\,1.2   &174 \,$\pm$\,26    &3.0 \,$\pm$\,0.5   &7.7 \,$\pm$\,1.2  &6.1 \,$\pm$\,1.1\\
71  &00:40:29.7 &65:53:50  &16.5\,$\pm$\,0.2  &1.50\,$\pm$\,0.04  &-104.8  &0.75\,$\pm$\,0.04  &4.1\,$\pm$\,0.1  &7.7 \,$\pm$\,0.4  &15.2 \,$\pm$\,1.3   &193 \,$\pm$\,29    &1.3 \,$\pm$\,0.2   &2.0 \,$\pm$\,0.3  &5.1 \,$\pm$\,1.0\\
72  &00:40:35.2 &65:53:19  &16.5\,$\pm$\,0.2  &0.93\,$\pm$\,0.03  &-103.6  &0.76\,$\pm$\,0.04  &3.3\,$\pm$\,0.1  &6.5 \,$\pm$\,0.3  &8.9  \,$\pm$\,0.8   &113 \,$\pm$\,17    &1.9 \,$\pm$\,0.4   &5.0 \,$\pm$\,0.8  &5.5 \,$\pm$\,1.1\\
73  &00:46:20.5 &64:38:05  &16.8\,$\pm$\,0.5  &1.03\,$\pm$\,0.05  &-102.3  &0.51\,$\pm$\,0.03  &4.6\,$\pm$\,0.1  &6.1 \,$\pm$\,0.3  &3.0  \,$\pm$\,0.7   &39.0\,$\pm$\,10.2  &0.5 \,$\pm$\,0.2   &1.3 \,$\pm$\,0.3  &7.9 \,$\pm$\,2.3\\
74  &00:46:22.3 &64:37:54  &16.8\,$\pm$\,0.5  &0.92\,$\pm$\,0.05  &-102.6  &0.50\,$\pm$\,0.02  &4.5\,$\pm$\,0.1  &5.8 \,$\pm$\,0.3  &13.3 \,$\pm$\,3.1   &175 \,$\pm$\,45    &3.1 \,$\pm$\,1.0   &8.0 \,$\pm$\,2.1  &1.5 \,$\pm$\,0.5\\
75  &00:47:06.0 &65:28:06  &16.6\,$\pm$\,0.2  &0.72\,$\pm$\,0.01  &-105.5  &0.75\,$\pm$\,0.03  &4.3\,$\pm$\,0.1  &8.2 \,$\pm$\,0.3  &19.2 \,$\pm$\,3.3   &245 \,$\pm$\,51    &7.0 \,$\pm$\,1.5   &23.3\,$\pm$\,4.8  &1.9 \,$\pm$\,0.5\\
76  &00:49:53.8 &65:23:57  &17.3\,$\pm$\,0.3  &1.48\,$\pm$\,0.04  &-107.3  &1.15\,$\pm$\,0.03  &7.6\,$\pm$\,0.1  &21.9\,$\pm$\,0.5  &130  \,$\pm$\,13    &1813\,$\pm$\,269   &12.3\,$\pm$\,1.9   &19.9\,$\pm$\,3.0  &1.3 \,$\pm$\,0.2\\
77  &00:50:00.3 &65:23:28  &17.3\,$\pm$\,0.3  &0.83\,$\pm$\,0.03  &-107.4  &0.53\,$\pm$\,0.03  &4.2\,$\pm$\,0.1  &5.7 \,$\pm$\,0.3  &7.2  \,$\pm$\,0.8   &100 \,$\pm$\,15    &2.2 \,$\pm$\,0.4   &6.2 \,$\pm$\,0.9  &2.7 \,$\pm$\,0.6\\
78  &01:37:48.8 &64:34:37  &17.4\,$\pm$\,0.5  &1.09\,$\pm$\,0.05  &-100.1  &0.87\,$\pm$\,0.04  &2.8\,$\pm$\,0.1  &6.2 \,$\pm$\,0.3  &24.0 \,$\pm$\,5.3   &339 \,$\pm$\,84    &4.2 \,$\pm$\,1.2   &9.3 \,$\pm$\,2.3  &2.8 \,$\pm$\,0.8\\
79  &01:37:43.8 &64:35:00  &17.4\,$\pm$\,0.5  &0.99\,$\pm$\,0.04  &-99.6   &0.90\,$\pm$\,0.05  &2.7\,$\pm$\,0.1  &6.0 \,$\pm$\,0.3  &21.2 \,$\pm$\,4.7   &298 \,$\pm$\,74    &4.5 \,$\pm$\,1.2   &10.9\,$\pm$\,2.7  &3.1 \,$\pm$\,0.9\\
80  &02:04:18.7 &63:14:34  &18.3\,$\pm$\,0.2  &0.81\,$\pm$\,0.02  &-102.5  &1.10\,$\pm$\,0.05  &4.0\,$\pm$\,0.1  &11.1\,$\pm$\,0.5  &42.3 \,$\pm$\,5.1   &675 \,$\pm$\,103   &15.2\,$\pm$\,2.7   &45.1\,$\pm$\,6.9  &1.7 \,$\pm$\,0.4\\
81  &02:05:09.0 &63:04:54  &18.6\,$\pm$\,0.4  &1.42\,$\pm$\,0.05  &-100.6  &1.00\,$\pm$\,0.04  &4.8\,$\pm$\,0.1  &12.1\,$\pm$\,0.5  &53.9 \,$\pm$\,7.2   &891 \,$\pm$\,146   &6.5 \,$\pm$\,1.2   &11.1\,$\pm$\,1.8  &1.8 \,$\pm$\,0.4\\
82  &02:20:46.6 &60:12:38  &17.9\,$\pm$\,0.2  &0.87\,$\pm$\,0.02  &-94.7   &0.56\,$\pm$\,0.02  &5.4\,$\pm$\,0.1  &7.7 \,$\pm$\,0.3  &13.8 \,$\pm$\,1.3   &209 \,$\pm$\,28    &4.1 \,$\pm$\,0.6   &11.3\,$\pm$\,1.5  &1.5 \,$\pm$\,0.3\\
83  &02:20:46.6 &60:12:54  &17.9\,$\pm$\,0.2  &0.94\,$\pm$\,0.02  &-95.0   &0.58\,$\pm$\,0.03  &5.1\,$\pm$\,0.1  &7.5 \,$\pm$\,0.3  &8.1  \,$\pm$\,0.8   &123 \,$\pm$\,17    &2.1 \,$\pm$\,0.3   &5.3 \,$\pm$\,0.7  &3.0 \,$\pm$\,0.6\\
84  &02:46:52.3 &58:18:46  &21.3\,$\pm$\,0.7  &1.17\,$\pm$\,0.06  &-99.0   &1.04\,$\pm$\,0.05  &4.4\,$\pm$\,0.1  &11.4\,$\pm$\,0.5  &37.6 \,$\pm$\,7.3   &884 \,$\pm$\,182   &9.6 \,$\pm$\,2.3   &19.6\,$\pm$\,4.1  &1.7 \,$\pm$\,0.4\\
85  &02:48:37.9 &58:28:43  &22.9\,$\pm$\,0.6  &2.26\,$\pm$\,0.08  &-103.9  &0.76\,$\pm$\,0.07  &5.8\,$\pm$\,0.2  &11.3\,$\pm$\,1.1  &85.5 \,$\pm$\,9.2   &2472\,$\pm$\,298   &7.2 \,$\pm$\,0.9   &7.6 \,$\pm$\,0.9  &0.6 \,$\pm$\,0.2\\
86  &02:48:42.3 &58:28:52  &22.9\,$\pm$\,0.6  &3.41\,$\pm$\,0.12  &-103.6  &1.09\,$\pm$\,0.03  &6.6\,$\pm$\,0.1  &18.2\,$\pm$\,0.5  &286  \,$\pm$\,31    &8250\,$\pm$\,993   &10.5\,$\pm$\,1.3   &7.4 \,$\pm$\,0.9  &0.6 \,$\pm$\,0.1\\
87  &02:48:34.1 &58:28:13  &22.9\,$\pm$\,0.6  &3.39\,$\pm$\,0.12  &-102.6  &1.56\,$\pm$\,0.07  &5.4\,$\pm$\,0.1  &21.3\,$\pm$\,0.9  &204  \,$\pm$\,22    &5900\,$\pm$\,710   &7.6 \,$\pm$\,0.9   &5.4 \,$\pm$\,0.7  &1.6 \,$\pm$\,0.3\\
88  &02:48:33.7 &58:27:45  &22.9\,$\pm$\,0.6  &1.29\,$\pm$\,0.05  &-101.9  &0.66\,$\pm$\,0.03  &4.9\,$\pm$\,0.1  &8.2 \,$\pm$\,0.3  &18.3 \,$\pm$\,2.0   &528 \,$\pm$\,64    &4.7 \,$\pm$\,0.7   &8.8 \,$\pm$\,1.1  &1.2 \,$\pm$\,0.2\\
89  &02:48:36.1 &58:28:30  &22.9\,$\pm$\,0.6  &1.98\,$\pm$\,0.07  &-102.3  &0.91\,$\pm$\,0.04  &5.1\,$\pm$\,0.1  &11.7\,$\pm$\,0.5  &48.9 \,$\pm$\,5.3   &1414\,$\pm$\,170   &5.3 \,$\pm$\,0.7   &6.5 \,$\pm$\,0.8  &1.3 \,$\pm$\,0.3\\
90  &02:48:37.1 &58:28:49  &22.9\,$\pm$\,0.6  &1.15\,$\pm$\,0.04  &-102.7  &0.60\,$\pm$\,0.03  &4.9\,$\pm$\,0.1  &7.5 \,$\pm$\,0.3  &18.1 \,$\pm$\,2.0   &524 \,$\pm$\,63    &5.9 \,$\pm$\,0.9   &12.3\,$\pm$\,1.5  &0.9 \,$\pm$\,0.2\\
91  &02:48:28.7 &58:23:23  &22.1\,$\pm$\,0.3  &1.22\,$\pm$\,0.03  &-102.1  &1.48\,$\pm$\,0.05  &6.0\,$\pm$\,0.1  &22.4\,$\pm$\,0.8  &112  \,$\pm$\,10    &2893\,$\pm$\,309   &28.8\,$\pm$\,3.5   &56.6\,$\pm$\,6.0  &1.1 \,$\pm$\,0.2\\
92  &02:48:29.3 &58:23:58  &22.1\,$\pm$\,0.3  &1.83\,$\pm$\,0.04  &-103.0  &0.90\,$\pm$\,0.03  &6.5\,$\pm$\,0.1  &14.9\,$\pm$\,0.4  &61.7 \,$\pm$\,5.5   &1599\,$\pm$\,171   &7.1 \,$\pm$\,0.8   &9.3 \,$\pm$\,1.0  &1.1 \,$\pm$\,0.2\\
93  &02:48:26.3 &58:23:47  &22.1\,$\pm$\,0.3  &0.76\,$\pm$\,0.02  &-102.4  &0.50\,$\pm$\,0.02  &5.6\,$\pm$\,0.1  &7.2 \,$\pm$\,0.3  &11.5 \,$\pm$\,1.1   &299 \,$\pm$\,32    &7.7 \,$\pm$\,1.1   &24.2\,$\pm$\,2.6  &0.7 \,$\pm$\,0.2\\
94  &02:48:29.6 &58:24:26  &22.1\,$\pm$\,0.3  &1.30\,$\pm$\,0.03  &-103.3  &0.71\,$\pm$\,0.03  &5.6\,$\pm$\,0.1  &10.1\,$\pm$\,0.4  &20.4 \,$\pm$\,1.8   &528 \,$\pm$\,56    &4.6 \,$\pm$\,0.6   &8.5 \,$\pm$\,0.9  &1.4 \,$\pm$\,0.2\\
95  &02:48:26.0 &58:22:45  &22.1\,$\pm$\,0.3  &1.22\,$\pm$\,0.03  &-102.5  &0.75\,$\pm$\,0.03  &4.8\,$\pm$\,0.1  &9.2 \,$\pm$\,0.4  &19.3 \,$\pm$\,1.7   &500 \,$\pm$\,53    &5.0 \,$\pm$\,0.6   &9.8 \,$\pm$\,1.1  &1.6 \,$\pm$\,0.3\\
96  &02:48:21.8 &58:22:39  &22.1\,$\pm$\,0.3  &1.21\,$\pm$\,0.03  &-102.8  &0.59\,$\pm$\,0.03  &5.0\,$\pm$\,0.1  &7.5 \,$\pm$\,0.3  &16.5 \,$\pm$\,1.5   &429 \,$\pm$\,46    &4.3 \,$\pm$\,0.5   &8.6 \,$\pm$\,0.9  &1.1 \,$\pm$\,0.2\\
97  &02:48:23.0 &58:23:41  &22.1\,$\pm$\,0.3  &1.17\,$\pm$\,0.03  &-102.3  &0.58\,$\pm$\,0.03  &5.1\,$\pm$\,0.1  &7.6 \,$\pm$\,0.3  &14.1 \,$\pm$\,1.3   &365 \,$\pm$\,39    &4.0 \,$\pm$\,0.5   &8.1 \,$\pm$\,0.9  &1.2 \,$\pm$\,0.2\\
98  &02:48:20.6 &58:23:27  &22.1\,$\pm$\,0.3  &1.44\,$\pm$\,0.03  &-101.9  &0.58\,$\pm$\,0.03  &4.6\,$\pm$\,0.1  &6.8 \,$\pm$\,0.3  &14.9 \,$\pm$\,1.4   &386 \,$\pm$\,41    &2.8 \,$\pm$\,0.3   &4.6 \,$\pm$\,0.5  &1.5 \,$\pm$\,0.3\\
99  &02:56:07.0 &57:25:41  &21.1\,$\pm$\,0.3  &1.14\,$\pm$\,0.03  &-96.7   &0.70\,$\pm$\,0.03  &7.8\,$\pm$\,0.1  &13.7\,$\pm$\,0.6  &27.1 \,$\pm$\,3.6   &620 \,$\pm$\,91    &7.1 \,$\pm$\,1.1   &14.9\,$\pm$\,2.2  &1.0 \,$\pm$\,0.2\\
100 &02:56:06.3 &57:25:37  &21.1\,$\pm$\,0.3  &1.00\,$\pm$\,0.02  &-97.6   &0.60\,$\pm$\,0.02  &6.9\,$\pm$\,0.1  &10.4\,$\pm$\,0.3  &11.8 \,$\pm$\,1.6   &270 \,$\pm$\,40    &4.0 \,$\pm$\,0.6   &9.6 \,$\pm$\,1.4  &1.5 \,$\pm$\,0.3\\
101 &03:31:33.3 &56:49:21  &17.7\,$\pm$\,0.4  &1.00\,$\pm$\,0.04  &-77.3   &1.00\,$\pm$\,0.04  &4.8\,$\pm$\,0.1  &12.0\,$\pm$\,0.5  &53.8 \,$\pm$\,8.5   &790 \,$\pm$\,150   &11.7\,$\pm$\,2.6   &28.1\,$\pm$\,5.3  &1.5 \,$\pm$\,0.4\\
102 &03:36:39.7 &56:43:04  &17.4\,$\pm$\,0.2  &0.97\,$\pm$\,0.03  &-75.1   &0.76\,$\pm$\,0.03  &4.5\,$\pm$\,0.1  &8.7 \,$\pm$\,0.4  &14.8 \,$\pm$\,2.5   &210 \,$\pm$\,41    &3.3 \,$\pm$\,0.7   &8.2 \,$\pm$\,1.6  &3.1 \,$\pm$\,0.7\\
103 &03:37:32.4 &55:07:25  &17.3\,$\pm$\,0.3  &0.94\,$\pm$\,0.03  &-74.0   &0.78\,$\pm$\,0.03  &8.1\,$\pm$\,0.1  &16.0\,$\pm$\,0.6  &12.8 \,$\pm$\,1.5   &179 \,$\pm$\,29    &3.0 \,$\pm$\,0.6   &7.7 \,$\pm$\,1.2  &3.7 \,$\pm$\,0.7\\
104 &03:37:32.1 &55:07:02  &17.3\,$\pm$\,0.3  &0.64\,$\pm$\,0.02  &-74.8   &0.70\,$\pm$\,0.02  &7.6\,$\pm$\,0.1  &13.3\,$\pm$\,0.4  &10.6 \,$\pm$\,1.3   &148 \,$\pm$\,24    &5.4 \,$\pm$\,1.2   &20.1\,$\pm$\,3.2  &2.4 \,$\pm$\,0.5\\
105 &03:37:32.9 &55:07:52  &17.3\,$\pm$\,0.3  &0.99\,$\pm$\,0.03  &-74.0   &0.68\,$\pm$\,0.02  &8.5\,$\pm$\,0.1  &14.6\,$\pm$\,0.4  &16.0 \,$\pm$\,1.9   &224 \,$\pm$\,36    &3.4 \,$\pm$\,0.6   &8.2 \,$\pm$\,1.3  &2.4 \,$\pm$\,0.5\\
106 &03:37:30.6 &55:07:26  &17.3\,$\pm$\,0.3  &0.83\,$\pm$\,0.03  &-74.9   &0.67\,$\pm$\,0.02  &7.6\,$\pm$\,0.1  &12.9\,$\pm$\,0.4  &9.5  \,$\pm$\,1.2   &134 \,$\pm$\,22    &2.9 \,$\pm$\,0.6   &8.3 \,$\pm$\,1.3  &3.2 \,$\pm$\,0.6\\
107 &03:37:34.2 &55:08:22  &17.3\,$\pm$\,0.3  &1.18\,$\pm$\,0.04  &-73.9   &0.72\,$\pm$\,0.02  &7.8\,$\pm$\,0.1  &14.1\,$\pm$\,0.3  &16.9 \,$\pm$\,2.0   &237 \,$\pm$\,38    &2.5 \,$\pm$\,0.5   &5.1 \,$\pm$\,0.8  &3.0 \,$\pm$\,0.6\\
108 &03:37:11.3 &53:36:38  &19.0\,$\pm$\,0.3  &0.94\,$\pm$\,0.03  &-77.5   &0.86\,$\pm$\,0.03  &6.7\,$\pm$\,0.1  &14.5\,$\pm$\,0.5  &30.7 \,$\pm$\,3.0   &535 \,$\pm$\,71    &9.0 \,$\pm$\,1.5   &22.9\,$\pm$\,3.0  &1.5 \,$\pm$\,0.3\\
109 &03:37:08.5 &53:36:52  &19.0\,$\pm$\,0.3  &0.81\,$\pm$\,0.02  &-77.4   &0.64\,$\pm$\,0.02  &6.1\,$\pm$\,0.1  &9.9 \,$\pm$\,0.3  &12.9 \,$\pm$\,1.3   &225 \,$\pm$\,30    &5.1 \,$\pm$\,0.8   &15.1\,$\pm$\,2.0  &1.7 \,$\pm$\,0.3\\
110 &03:37:11.9 &53:37:09  &19.0\,$\pm$\,0.3  &0.88\,$\pm$\,0.03  &-77.8   &0.64\,$\pm$\,0.03  &6.1\,$\pm$\,0.1  &9.9 \,$\pm$\,0.4  &12.7 \,$\pm$\,1.3   &220 \,$\pm$\,29    &4.2 \,$\pm$\,0.7   &11.5\,$\pm$\,1.5  &1.9 \,$\pm$\,0.4\\
111 &03:37:10.5 &53:37:02  &19.0\,$\pm$\,0.3  &0.58\,$\pm$\,0.02  &-77.1   &0.58\,$\pm$\,0.03  &5.6\,$\pm$\,0.1  &8.3 \,$\pm$\,0.4  &4.0  \,$\pm$\,0.4   &69.5\,$\pm$\,9.2   &3.1 \,$\pm$\,0.7   &12.7\,$\pm$\,1.7  &3.3 \,$\pm$\,0.6\\
112 &03:37:14.5 &53:37:21  &19.0\,$\pm$\,0.3  &0.99\,$\pm$\,0.03  &-78.3   &0.58\,$\pm$\,0.03  &5.4\,$\pm$\,0.1  &8.0 \,$\pm$\,0.3  &8.4  \,$\pm$\,0.9   &145 \,$\pm$\,19    &2.2 \,$\pm$\,0.4   &5.3 \,$\pm$\,0.7  &2.7 \,$\pm$\,0.5\\
\hline
\end{tabular}
\label{table:counts}
\end{table*}
\endgroup

\newpage
\onecolumn
\section{Fitted results of the scaling relations}
\label{Sect:Fitted results of the scaling relations}

\begin{table*}[h]
\caption{Fitted results of the scaling relations.}
\centering
\begin{tabular}
{lccccccccc}
\hline\hline
Type &Number &$m$ &$v_0$ &$r_{\rm v}$ &$p_{\rm v}$ &$q$ &$M_0$ &$r_{\rm M}$ &$p_{\rm M}$\\
\hline
Perseus arm   &12464 &0.42\,$\pm$\,0.01 &0.38\,$\pm$\,0.01 &0.62 &<\,10$^{-3}$ &2.41\,$\pm$\,0.01 &19.3\,$\pm$\,0.2   &0.96 &<\,10$^{-3}$\\
Outer arm     &5144  &0.54\,$\pm$\,0.01 &0.25\,$\pm$\,0.01 &0.68 &<\,10$^{-3}$ &2.37\,$\pm$\,0.02 &20.8\,$\pm$\,0.5   &0.95 &<\,10$^{-3}$\\  
OSC arm       &428   &0.63\,$\pm$\,0.04 &0.18\,$\pm$\,0.02 &0.64 &<\,10$^{-3}$ &2.18\,$\pm$\,0.07 &30.9\,$\pm$\,4.3   &0.86 &<\,10$^{-3}$\\
Galactic edge &112   &0.36\,$\pm$\,0.10 &0.69\,$\pm$\,0.03 &0.34 &<\,10$^{-3}$ &2.18\,$\pm$\,0.26 &196\,$\pm$\,17     &0.63 &<\,10$^{-3}$\\
\hline
Compact       &81    &0.31\,$\pm$\,0.11 &0.73\,$\pm$\,0.03 &0.32 &0.004        &2.27\,$\pm$\,0.29 &210\,$\pm$\,23     &0.66 &<\,10$^{-3}$\\
Intermediate  &28    &0.67\,$\pm$\,0.28 &0.59\,$\pm$\,0.05 &0.46 &0.013        &1.85\,$\pm$\,0.63 &164\,$\pm$\,30     &0.55 &0.003       \\
\hline
With SF       &73    &0.36\,$\pm$\,0.11 &0.76\,$\pm$\,0.03 &0.39 &<\,10$^{-3}$ &2.46\,$\pm$\,0.31 &216\,$\pm$\,23     &0.70 &<\,10$^{-3}$\\
Possible SF   &17    &0.38\,$\pm$\,0.24 &0.66\,$\pm$\,0.05 &0.44 &0.075        &1.75\,$\pm$\,0.83 &177\,$\pm$\,47     &0.55 &0.023       \\
Without SF    &22    &0.10\,$\pm$\,0.34 &0.56\,$\pm$\,0.06 &0.06 &0.782        &0.46\,$\pm$\,0.52 &166\,$\pm$\,24     &0.22 &0.323       \\
\hline
\end{tabular}
\label{table:Fit results for the scaling relations}
\tablefoot{Column 1: Types of molecular clouds. Molecular clouds are located in the Perseus arm, Outer arm, OSC arm,
and Galactic edge (\emph{this work}) (see Figs.\,\ref{fig:size-velocity dispersion} and \ref{fig:Reff-Mlum}), which are 
classified as compact or intermediate, with star-forming activity, possible star-forming activity, or without star-forming 
activity (see Figs.\,\ref{fig:size-linewidth-class} and \ref{fig:size-mass-class}). Column 2: The number of samples of each 
fitted cloud type. Columns 3--6: The fitted parameters of the $\sigma_v$--$R$ relation, 
including the power-law index $m$, the normalization coefficient $v_0$, the Pearson correlation coefficient $r_{\rm v}$, 
and the statistical significance $p_{\rm v}$. Columns 7--10: The fitted parameters of the $M$--$R$ relation, 
including the power-law index $q$, the normalization coefficient $M_0$, the Pearson correlation coefficient $r_{\rm M}$, 
and the statistical significance $p_{\rm M}$. Power-law fitting and the Pearson correlation coefficients 
(with statistical significance $p$-values) were calculated using the {\tt linmix} module from \citet{Kelly2007} and 
the {\tt pearsonr} function of the {\tt Python SciPy} package, respectively. The fitting was performed in log--log space 
using a linear regression method with uncertainties on both axes included.} 
\end{table*}

\end{appendix}
\end{document}